\documentclass[aps,floatfix,prd,showpacs ]{revtex4}
\usepackage{graphicx}
\usepackage{dcolumn}
\usepackage{bm}
\usepackage{enumerate}
\usepackage{tipa}

\usepackage{amsmath}
\usepackage{slashed}
\usepackage{graphicx}
\usepackage{subfigure}
\usepackage{float}
\usepackage{epstopdf}
\usepackage{color}
 \usepackage{feyn}
\definecolor{darkgreen}{rgb}{0,0.5,0}
\definecolor{purple}{rgb}{0.5,0,0.5}
\definecolor{nblue}{rgb}{0.0,0.0,0.50}
\definecolor{scarlet}{rgb}{1.0,0.2,0}
\usepackage[colorlinks=true,linkcolor=purple,citecolor=purple,urlcolor=blue]{hyperref}
\begin{document}

\def\no{\noindent}
\def\non{\nonumber\\}
\def\epsp#1#2{\varepsilon_{#1}\cdot p_{#2}}
\def\be{\begin{equation}}
\def\ee{\end{equation}}
\def\bea{\begin{eqnarray}}
\def\eea{\end{eqnarray}}
\def\bear{\begin{eqnarray}}
\def\ear{\end{eqnarray}}
\def\bei{\begin{itemize}}
\def\eei{\end{itemize}}
\def\bee{\begin{enumerate}}
\def\eee{\end{enumerate}}
\def\noN{\nonumber}
\def\ccr{\nonumber\\}
\def\eqa{\!&=&\!}
\def\Eins{{\mathchoice {\rm 1\mskip-4mu l} {\rm 1\mskip-4mu l}
{\rm 1\mskip-4.5mu l} {\rm 1\mskip-5mu l}}}
\def\Z{{\mathchoice {\hbox{$\sf\textstyle Z\kern-0.4em Z$}}
{\hbox{$\sf\textstyle Z\kern-0.4em Z$}}
{\hbox{$\sf\scriptstyle Z\kern-0.3em Z$}}
{\hbox{$\sf\scriptscriptstyle Z\kern-0.2em Z$}}}}
\def\e{\,{\rm e}}
\def\veps#1{\varepsilon_{#1}}
\def\eps#1{\epsilon_{#1}}
\def\kk#1#2{k_{#1}\cdot k_{#2}}
\def\pp#1#2{p_{#1}\cdot p_{#2}}
\def\intT{\int_0^\infty dT\, {\rm e}^{-m^2T}}
\def\matD{\mathcal{D}}
\def\t #1{\tau_{#1}}
\def\eTx{{\rm e}^{-\frac{1}{4}\int_0^Td\tau\dot{x}^2}}
\def\eTq{{\rm e}^{-\frac{1}{4}\int_0^Td\tau\dot{q}^2}}
\def\vaeps{\varepsilon}
\def\ddel{{}^\bullet\! \Delta}
\def\deld{\Delta^{\hskip -.5mm \bullet}}
\def\dddel{{}^{\bullet \bullet} \! \Delta}
\def\ddeld{{}^{\bullet}\! \Delta^{\hskip -.5mm \bullet}}
\def\deldd{\Delta^{\hskip -.5mm \bullet \bullet}}
\def\epsk#1#2{\varepsilon_{#1}\cdot k_{#2}}
\def\epseps#1#2{\varepsilon_{#1}\cdot\varepsilon_{#2}}
\def\Gd{\dot{G}}
\def\Gdd{\ddot{G}}

\title{Multiphoton amplitudes and generalized LKF transformation in Scalar QED}
\author{Naser Ahmadiniaz$^{a,b,d}$, Adnan Bashir$^a$ and Christian Schubert$^{a,c}$}
\affiliation{$^a$ Instituto de F{{\'\i}}sica y Matem\'aticas, Universidad Michoacana de San Nicol\'as de Hidalgo\\
Apdo. Postal 2-82, C.P. 58040, Morelia, Michoacan, Mexico\\
$^b$ Facultad de Ciencias en F{{\'\i}}sica y Matem\'aticas, Universidad Aut\'onoma de Chiapas, Ciudad Universitaria, Tuxtla Guti\'errez 29050, \\
Tuxtla Guti\'errez, M\'exico\\
$^c$
Max-Planck-Institut f\"ur Gravitationsphysik\\
 Albert-Einstein-Institut\\
M\"uhlenberg 1, D-14476 Potsdam, Germany\\
$^d$
Center for Relativistic Laser Science,\\
 Institute for Basic Science,\\
 Gwangju 500-712, Korea
}

\date{\today}

\begin{abstract}
We apply the worldline formalism to amplitudes in scalar quantum
electrodynamics (QED) involving open scalar lines, with an
emphasis on their non-perturbative gauge dependence. At the
tree-level, we study the scalar propagator interacting with any
number of photons in configuration space as well as in momentum
space. At one-loop we rederive, in an efficient way, the off-shell
vertex in an arbitrary dimension and any covariant gauge.
Generalizing the Landau-Khalatnikov-Fradkin transformation (LKFT)
for the non-perturbative propagator, we find simple
non-perturbative transformation rules for arbitrary $x$-space
amplitudes under a change of the covariant gauge parameter in
terms of conformal cross ratios.

\end{abstract}

\maketitle

\section{Introduction}
\label{intro}
\renewcommand{\theequation}{1.\arabic{equation}}
\setcounter{equation}{0}

Unravelling the non-perturbative structure of Green functions in
gauge field theories has been a challenging task. A deep
understanding of the emergent phenomena of confinement and
dynamical chiral symmetry breaking in quantum chromodynamics (QCD)
can only be achieved through the outcome of this endeavor.
However, leaving aside the intricacies of a non abelian gauge
field theory, there is a lot one can learn from the relatively
simpler cases of spinor and scalar quantum electrodynamics (QED).

Perturbation theory and gauge covariance properties of Green
functions have served as guiding principles towards our knowledge
of their non-perturbative counterparts. Interaction vertices are
naturally a focus of study in this context. A systematic study of
the 3-point electron-photon vertex in spinor QED was initiated
more than three decades ago by Ball and Chiu~\cite{bach-80}. They
decompose the vertex into longitudinal and transverse parts, where
the former satisfies the Ward-Fradkin-Green-Takahashi identity
(WFGTI)~\cite{WGTI-1950}. They provide a set of eight basis
vectors to write out the transverse vertex and calculate it at the
one-loop level for off-shell external legs in 4-dimensions in the
Feynman gauge. Their choice of the transverse basis guarantees
that each of the corresponding coefficients is independent of any
unwanted kinematic singularities. 

Similarly, the one loop
electron-photon vertex in Yennie-Fried gauge was evaluated
in~\cite{adkins-94}. Later, the work of Ball and Chiu was extended
to arbitrary covariant gauges by K\textsci z\textsci lers\"u {\em
et. al.}~\cite{Pennington-95}. However, this work shows that a
slight change in the transverse basis is required to ensure the
absence of kinematic singularities for each and every coefficient
in an arbitrary covariant gauge. For the massive and massless 3-point
vertex in 3-dimensional spinor QED (${\rm QED}_3$), the results in
arbitrary covariant gauges were obtained
in~\cite{adnan1,adnan2,adnan3}. The super renormalizability of
${\rm QED}_3$ implies that the vertex has no ultraviolet
divergences. ${\rm QED}_3$ thus provides a neater laboratory to
explore the effects of the electron-photon vertex on dynamical
chiral symmetry breaking and confinement.

The QED vertex is useful not only by itself, but it also serves as
a simple model for its subsequent extension to the more
complicated non-abelian case of QCD. Due to the identical Dirac nature
of electrons and quarks, the quark-gluon vertex has the same
number of basis vectors in its general decomposition, twelve, as the
electron-photon one.

The difference lies in the fact that it is now the Slavnov-Taylor
identity (STI)~\cite{Slavnov-Taylor-1971} which extracts out the
longitudinal part, still leaving eight basis vectors for expanding
out the transverse vertex. There have been several works on the
calculation of the quark-gluon vertex at one-loop and beyond in
different kinematic regimes of momenta~\cite{Nowak-1986},
relating the symmetric point vertex to the running coupling in
QCD. The systematic generalization of the ${\rm QED_3}$-point
vertex to the study of QCD can be attributed to Ball and
Chiu~\cite{bach-qcd}. Identically to the case of QED, they
calculate the quark-gluon vertex to the one loop order in the
Feynman gauge and project it onto the basis they proposed earlier.
Employing the modified basis of~\cite{Pennington-95}, Davydychev
{\em et. al.}~\cite{davossa-2000}, evaluate the one-loop
quark-gluon vertex in arbitrary gauge $\xi$ and spacetime
dimensions $D$ in an $SU(N)$ gauge field theory. An appropriate
choice of $\xi$, $D$ and the color factors reproduces the results
of earlier QED and QCD studies, just as one would expect. It is
important to note that the knowledge of the 3-point interactions
in an arbitrary covariant gauge is a crucial guiding principle to
pinpoint the transverse vertex, which remains undetermined through
the WFGTI or STI. {\em Can the knowledge of a 3-point vertex in
one gauge lead us to know what is it in any other covariant gauge
without having to redo the calculation ab initio?} This is the
kind of question we address in this article.

Our focus of attention is an even simpler gauge theory, namely
scalar QED. Due to the absence of Dirac matrices, it only requires
two basis vectors for its most general description. The
longitudinal one is fixed by the WFGTI. The transverse part of the
one loop vertex was calculated by Ball and Chiu~\cite{bach-80}
in the Feynman gauge. However, the study of the gauge dependence
of the Green functions and their connection to the lower point
functions require their knowledge in an arbitrary covariant gauge.
This study was carried out by Bashir {\em et. al.}
in~\cite{adnan-4,adnan-5}. In the present work, adopting the
string-inspired worldline
formalism~\cite{feyn-sqed,feyn-qed,berkos,strassler-92}, we set
out to {\em construct this vertex in an arbitrary covariant gauge
based on its knowledge solely in one gauge}.

On more general grounds, the gauge dependence of Green functions
was formally studied by Landau and Khalatnikov~\cite{LK}, and
independently by Fradkin~\cite{fradkin}. They derived a series of
transformations, dubbed as Landau-Khalatnikov-Fradkin
transformations (LKFT), which encapsulate how the Green functions
transform in a specific manner under a variation of gauge. The
LKFT were later re-derived by Johnson and Zumino through the use
of functional methods~\cite{john-zum}. These transformations are
non-perturbative, and written in coordinate space; obtaining an
analytical counterpart for them in momentum space is not a simple
task. The LKFT can not only be used to change from one covariant
gauge to another at a fixed loop level, but also to predict
higher-loop terms from lower-loop ones. However, those predicted
terms will all be gauge parameter dependent~\cite{LKFT-Loops}.

In the standard second-quantized formalism, the LKFT for the
fermion propagator is relatively easier to investigate than for
the scalar one. Studies in massless scalar and spinor QED, as well
as in QCD (under certain conditions), demonstrate that the
wavefunction renormalization  has a multiplicatively
renormalizable form of a power law in
4-dimensions~\cite{LKFT-Loops,Aslam-2015}. In the quenched QED
gap equation for the electron propagator, this solution can be
reproduced only with an appropriate choice of the electron-photon
3-point vertex. It is well known that the longitudinal Ball-Chiu
vertex is not sufficient to ensure the LKFT law~\cite{curtpenn}.
Since that realization, there have been a series of works which
construct the electron-photon vertex implementing the
multiplicative renormalizabilty of the electron
propagator~\cite{curtpenn, roberts,adnpenn,pennkiz}.
In~\cite{adnanroberts}, this LKFT law was implemented for the
fermion propagator, and it simultaneously ensures the gauge
invariance of the critical coupling above which chiral symmetry is
dynamically broken. It also provides an infrared enhanced
anomalous magnetic moment for quarks, advocated
in~\cite{enhancedmm}.

All the above efforts are about constructing a 3-point vertex
which ensures the LKFT of the fermion
propagator in its gap equation. However, the 3-point vertices have
their own transformation laws that were derived
in~\cite{LK,fradkin,john-zum}. {\em In this article, the local
gauge transformation of the 3-point scalar QED vertex is
addressed through the worldline approach.}

In the present article, several of the above issues will be
studied for the scalar QED case. We have three main motivations
and objectives in mind: firstly, the multi-photon generalizations
of Compton scattering are becoming important these days for laser
physics, see for example~\cite{ilderton} and~\cite{di piazza}.
Secondly, the computation of off-shell form factors for $n$-point
Green functions in scalar QED can provide a simple yet non-trivial
and insightful starting point to later go on to study spinor QED
or QCD. And thirdly, we look for efficient ways of transforming a
Green function from one covariant gauge to another, independently
for external and internal photons. In fact, the second and third
objectives are interrelated as it is usually not sufficient to
know form factors in just one gauge. In order to establish the
gauge invariance of associated physical observables and to look
for relations between different Green functions, their knowledge
in different covariant gauges can be of valuable help. This
observation is intricately related to the LKFT~\cite{LK,fradkin}.

Our method of choice here is not the standard second-quantized formalism,
but the  string-inspired worldline formalism.
Prior to embarking upon our main discussion, let us
provide a brief summary of this not so well-known formalism
(see \cite{22,41} for reviews).

One of the main reasons for studying  string theory is the fact
that it provides us with an efficient mathematical framework that
transcends quantum field theory, but reduces to it in the limit of
infinite string tension. A systematic investigation of this limit
led Bern and Kosower in 1991~\cite{berkos} to a novel and
efficient way to compute gauge theory amplitudes. In particular,
they obtained a compact generating function for the one-loop (on-shell) $N$ -
gluon amplitudes, known as the {\it Bern-Kosower master formula},
and they applied it to a first calculation of the five-gluon
amplitudes~\cite{bediko5glu}. 

Later, Strassler~\cite{strassler-92} showed that many of their results
can be obtained more straightforwardly using a representation of
the S-matrix in terms of first-quantized particle path integrals,
invented for the QED case by Feynman in
1950~\cite{feyn-sqed,feyn-qed}. This ``string-inspired worldline
formalism'' is manifestly off-shell, and uses string theory only
as a guiding principle. In contrast to the standard Feynman
diagrammatic approach, the main advantages which one can hope to
obtain by employing this formalism, are the following:

\begin{enumerate}

\item
Effectively the loop momenta have already been integrated out, which reduces the number
of possible kinematic invariants from the beginning.

\item In favorable cases, it allows one to derive compact master
formulas that contain the information on large numbers of Feynman
diagrams~\cite{10,15,dashsu,41,100,ahbaco-colorful}.

\item In gauge theory, it is frequently possible to achieve
manifest gauge invariance already {\em at the integrand
level}~\cite{strassler-92,26,92,98}.

\item The method treats spin in a more uniform manner; in
particular, calculations of amplitudes in spinor QED usually yield
the corresponding results for scalar QED as a
spin-off~\cite{strassler-92,26}.

\item It is generally easier than in the standard approach to
incorporate constant external
fields~\cite{shaisultanov,17,18,24,40}.

\end{enumerate}

To give a recent example, two of the present authors have used the
formalism to recalculate the off-shell three-gluon
vertex~\cite{92}, recuperating the form-factor decomposition
originally proposed by Ball and Chiu~\cite{ball-chiu-3gluon}, in
a way that not only reduced the amount of algebra significantly,
but also made the analysis of the Ward-identities unnecessary.
Moreover, it unified the scalar, spinor and gluon loop cases. The
superior efficiency of the method becomes even more conspicuous at
the four-gluon level~\cite{98,4-gluon-main}.

For scalar QED,  as early as 1996 Daikouji {\em et. al.}~ \cite{dashsu} showed
how to apply the string-inspired worldline formalism to arbitrary
amplitudes, albeit only in momentum space. However, for our present study of the
non-perturbative gauge-dependence, it will be essential to work in $x$-space
as well as in momentum space. 

The structure of this paper is as follows. In sections II and III we give short summaries of the 
worldline formalism and the LKFT, respectively. In
section IV, we derive our master formula for the scalar propagator
dressed by an arbitrary number of photons, both in configuration
and in momentum space. We then use this tree-level formula in
section V to construct, by sewing, the one-loop corrections to the
scalar propagator and to the photon-scalar 3-point vertex, in
Feynman gauge and for arbitrary space-time dimension. In section
VI, we derive our central result, which is a generalization of the
LKFT to arbitrary $x$-space amplitudes in scalar QED. The argument
is based on the fact that changes of the
covariant gauge parameter correspond to total derivative terms
in the worldline integrands. This formula is non-perturbative. In
section VII, we take up an example to show how it works at the
perturbative level. In section VIII, we return to the one-loop
propagator and vertex in momentum space, and show how to obtain
their form in an arbitrary covariant gauge from the one in Feynman
gauge in an efficient way by identifying all difference terms as
total derivative terms. We then compare our results with the ones
obtained in~\cite{adnan-4} and observe complete agreement. In
section~\ref{conclusion}, we summarize our results and discuss
possible generalizations. 
There are two appendices: in Appendix \ref{app-integrals} we give explicit
results for some of the integrals appearing in the calculation of the one-loop
propagator and vertex. Some details of the calculation of the vertex 
have been relegated to Appendix \ref{app-details}. 

We work with Euclidean conventions throughout. 

\section{The worldline formalism in scalar QED}
\label{wl formalism}
\renewcommand{\theequation}{2.\arabic{equation}}
\setcounter{equation}{0}

In this section, we discuss our method, which is based on the
worldline formalism, developed by Feynman for scalar
QED~\cite{feyn-sqed} and spinor QED~\cite{feyn-qed}, as an
extension of  his more standard path integral formalism for
non-relativistic quantum mechanics.

Although the formalism applies to arbitrary amplitudes in scalar
QED, the essential features of the formalism can be seen already
from the case of the quenched scalar propagator. Feynman's path
integral representation~\cite{feyn-sqed} of the quenched
propagator of a scalar particle of mass $m$, which propagates from
point $x'$ to $x$ in the presence of a background $U(1)$ gauge
field $A$, is

 \bear
 \Gamma[x,x']&=&
 \int_0^\infty dT\, {\rm e}^{-m^2T}\,
  \int_{x(0)=x'}^{x(T)=x}\matD
 x(\tau)\, {\rm e}^{-S_0 - S_e - S_i} \,,
 \label{scalar-propagator}
 \ear
 where
 \bear
 S_ 0 &=&  \int_0^Td\tau \frac{1}{4}\dot{x}^2\,, \nonumber\\
 S_e &=& ie\int_0^Td\tau\, \dot{x}\cdot A(x(\tau))\,, \nonumber\\
 S_i &=& \frac{e^2}{2}
\int_0^Td\tau_1\int_0^Td\tau_2\, \dot{x}^{\mu}_1 D_{\mu\nu}(x_1-x_2)\dot{x}^{\nu}_2\,.
\nonumber\\
\label{defS}
 \ear
 Here, $S_0$ describes the free propagation,
$S_e$ the interaction of the scalar with the external field, and
$S_i$ virtual photons exchanged along the scalar's trajectory. We
abbreviate $x(\tau_i) \equiv x_i$ etc. $D_{\mu\nu}$ is the $x$ -
space photon propagator in $D$ dimensions. In an arbitrary
covariant gauge, it is given by

\bear
D_{\mu\nu}(x) =
\frac{1}{4\pi^{\frac{D}{2}}}
\Big\{\frac{1+\xi}{2}\Gamma\Big(\frac{D}{2}-1\Big)\frac{\delta_{\mu\nu}}{{x^2}^{\frac{D}{2}-1}}+(1-\xi)\Gamma\Big(\frac{D}{2}\Big)
\frac{x_{\mu}x_{\nu}}{{x^2}^{\frac{D}{2}}}\Big\}\,.
\label{covariant-gauge-trans} \ear Here $\xi=1$ corresponds to
Feynman gauge and $\xi=0$ to Landau gauge.

\begin{figure}[h]
  \centering
    \includegraphics[width=0.85\textwidth]{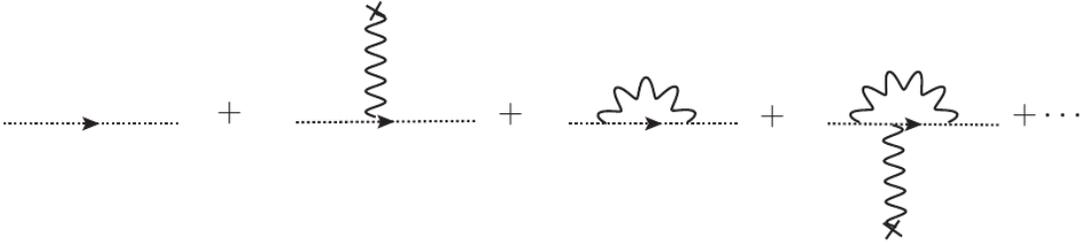}
\caption{Sum of Feynman diagrams for the dressed quenched propagator represented by a single path
integral.}
\label{fig-quenched_prop}
\end{figure}
\noindent

The expansion of the exponentials of the interaction terms $S_e$
and $S_i$ generates the Feynman diagrams depicted in FIG.
\ref{fig-quenched_prop}. The external legs represent interactions
with the field $A(x)$, and are converted into momentum-space
photons by choosing $A(x)$ as a sum of plane waves,
 \bear
A^{\mu}(x) = \sum_{i=1}^N \varepsilon_i^{\mu}\,\e^{ik_i\cdot x}\,.
\label{decompA}
\ear
 Each external photon then gets represented by a {\it vertex
 operator}.
 \bear
V_{\rm scal}^A[k,\vaeps] \equiv
\veps\mu\int_0^Td\tau\,\dot{x}^\mu(\tau)\,{\rm e}^{ik\cdot
x(\tau)} \,. \label{defvertop}
 \ear
According to our convention, external photon momenta are ingoing.
Note that the integrand in $S_i$, as defined in Eq.~(\ref{defS}),
may also be written as
 \bear
 \frac{1}{4\pi^{\frac{D}{2}}}
 \biggl\lbrack
 \Gamma\Big(\frac{D}{2}-1\Big)\frac{\dot{x}_1\cdot\dot{x}_2}{[(x_1-x_2)^2]^{\frac{D}{2}-1}}-\frac{1-\xi}{4}
 \Gamma\Big(\frac{D}{2}-2\Big)\frac{\partial}{\partial\tau_1}\frac{\partial}{\partial\tau_2}[(x_1-x_2)^2]^{2-\frac{D}{2}}
 \biggr\rbrack
 \,,
 \label{gauge-D}
 \ear
which shows, already at this level, that a change of the covariant
gauge parameter creates only a total derivative term.

The path integral is computed by splitting $x^\mu(\tau)$ into a
``background'' part $x^{\mu}_{\rm bg}(\tau)$, which encodes the
boundary conditions, and a fluctuation  part $q^\mu(\tau)$, which
has Dirichlet boundary conditions at the endpoints $\tau =0, T$:
 \bear x(\tau)&=&x_{\rm
bg}(\tau)+q(\tau)\,,\non x_{\rm
bg}(\tau)&=&x'+\frac{(x-x')\tau}{T}\,,\non
\dot{x}(\tau)&=&\frac{x-x'}{T}+\dot{q}(\tau)\,,\non
 q(0)&=&q(T)=0\,.
 \nonumber\\
 \label{split}
\ear The path integral over the fluctuation variable $q(\tau)$ is
gaussian, except for the denominators of the photon exchange terms
$S_i$. A fully gaussian representation is achieved by the further
introduction of a photon proper-time~\cite{15}, rewriting
\begin{equation}
{\Gamma(\lambda )\over {4{\pi}^{\lambda +1}}
{\Bigl({[x(\tau_a) - x(\tau_b)]}^2\Bigr)}^{\lambda}}
=\int_0^{\infty} d\bar T
{(4\pi \bar T)}^{-{D\over 2}}
{\rm exp}\Biggl
[-{{\Bigl(x(\tau_a)-x(\tau_b)\Bigr)}^2\over 4\bar T}
\Biggr ]\,.
\label{intpropchap8}
\end{equation}
\no
The calculation of the path integral then requires only the knowledge of the
free path integral normalization, which is
\bear
\int\mathcal{D}q(\tau)\,{\rm e}^{-\int_0^Td\tau\frac{1}{4}\dot{q}^2} = (4\pi T)^{-\frac{D}{2}}\,,
\ear
and of the two-point correlator, given by \cite{mckeon:ap224,basvan-book}
\bear \langle
q^\mu(\tau_1)q^\nu(\tau_2)\rangle&=&-2\delta^{\mu\nu}\Delta(\tau_1,\tau_2)\,,
\label{wick} \ear with the worldline Green function
$\Delta(\tau_i,\tau_j)$,
\bear
\Delta(\tau_1,\tau_2)&=&\frac{\tau_1\tau_2}{T}+\frac{\vert\tau_1-\tau_2\vert}{2}-\frac{\tau_1+\tau_2}{2}\,.
\label{defDelta} \ear We note that this Green function has a
nontrivial coincidence limit
\bear
\Delta(\tau,\tau)&=&\frac{\tau^2}{T}-\tau\,,
\label{coinDelta}
\ear
and we will also need its following derivatives:
\bear \ddel(\tau_1,\tau_2)&=&\frac{\tau_2}{T}+\frac{1}{2}{\rm
sign}(\tau_1-\tau_2)-\frac{1}{2}\,,\non
\deld(\tau_1,\tau_2)&=&\frac{\tau_1}{T}-\frac{1}{2}{\rm
sign}(\tau_1-\tau_2)-\frac{1}{2}\,,\non
\ddeld(\tau_1,\tau_2)&=&\frac{1}{T}-\delta(\tau_1-\tau_2)\,.
\nonumber\\
\label{derDelta}
 \ear
 Here we follow the notation~\cite{basvan-book} that left and right dots indicate derivatives
with respect to the first and the second argument, respectively. Note that the
mixed derivative $\ddeld(\tau_1,\tau_2)$ contains a delta function
which brings together two photon legs; this is how the seagull
vertex arises in the worldline formalism. 

In the simplest case,
for the free scalar propagator, we thus get the following standard
proper-time representation in $D$ dimensions:
\bear
\Gamma_{\rm free}[x,x']=\int_0^\infty dT\,{\rm e}^{-m^2T}\,
 (4\pi T)^{-\frac{D}{2}}
{\rm e}^{-\frac{1}{4T}(x-x')^2}\,.
 \ear

\section{Landau-Khalatnikov-Fradkin transformations}
\label{lkft}
\renewcommand{\theequation}{3.\arabic{equation}}
\setcounter{equation}{0}

LKFTs are rules which transform Green functions in a specific
manner from one covariant gauge to another. The set of LKFT are
non-perturbative in nature and we have already discussed the
history of their derivation and applications in
section~\ref{intro}. It is clear from the pioneering works of
Landau, Khalatnikov and Fradkin~\cite{LK,fradkin}, that these
transformations work similarly for spinor and scalar QED. They
give explicit rules with closed formulas to all orders in
coordinate space for the two- and three-point functions. The
momentum space treatment of the LKFT was carried out
perturbatively at one- or two-loop orders in~\cite{LKFT-Loops}. 
Their non-perturbative implementation in momentum space was
performed numerically in ${\rm QED_3}$ to establish the gauge
invariance of chiral symmetry breaking and
confinement~\cite{nonp-LKFT1,nonp-LKFT2}.

Let us look at the derivation of the LKFT for the two-point
propagator, following~\cite{LK,fradkin}. The photon propagator in
the coordinate space can be written as 

\bear
D_{\mu\nu}(x,f)=D_{\mu\nu}(x,0)+\partial_\mu
 \partial_\nu f_D(x)\,, \label{photon-propagator}
 \ear
where $f_D(x)$ is some function which corresponds to a particular
gauge fixing procedure. In covariant gauge for $D$ space-time
dimension, its explicit form is
 \bear f_D(x)=-i\xi e^2\mu^{4-D}\int
 \frac{d^Dk}{(2\pi)^D}\frac{{\rm e}^{-ik\cdot x}}{k^4}\,, \label{delta}
 \ear
where $\mu$ is the usual mass scale, introduced to ensure 
that the coupling $e$ remains dimensionless in every dimension
$D$. By means of dimensional regularization, this
integral can be evaluated to obtain:
  \bear f_D(x)=-\frac{ie^2\xi}{16{\pi}^{\frac{D}{2}}}(\mu
 x)^{4-D}\Gamma\Big(\frac{D}{2}-2\Big)\,. \label{delta-lkft}
 \ear
In~\cite{adnan3}, LKFTs were used to obtain the covariant gauge
representation of the fermion propagator, starting from its
knowledge in the Landau gauge for $D=3$ and $D=4$. For the three
dimensional case, using $\alpha = e^2/(4 \pi)$, one gets
 \bear
 f_3(x)=-\frac{i\alpha\xi x}{2}\,,
 \ear
which leads to the following fermion propagator in an arbitrary
gauge:
 \bear S_F(x;\xi)=S_F(x;0)\, {\rm e}^{-(\alpha\xi/2)x}\,.
 \ear
Fourier transforming to momentum space, they recover the known
results for the wave function renormalization $F(p;\xi)$ and the mass
function $\mathcal{M}(p;\xi)$ at the one loop order up to a term
proportional to $\alpha\xi^0$, see~\cite{adnan3}, as permitted by
the structure of the LKFT.

We now look at the 4-dimensional case. By expanding
Eq.~(\ref{delta-lkft}) around $D=4-\epsilon$, and using
 \bear
 \Gamma\Big(-\frac{\epsilon}{2}\Big)&=&-\frac{2}{\epsilon}-\gamma+\mathcal{O}(\epsilon)\,,
 \ear
we obtain
 \bear f_4(x)=i\frac{\xi
e^2}{16\pi^{2}}\Big[\frac{2}{\epsilon}+\gamma+\ln \pi + 2\ln (\mu
x)+\mathcal{O}(\epsilon)\Big]\,.
 \ear
Thus in the four-dimensional case one has to be more careful.
Note that in the term
proportional to $\ln x$ one cannot simply put $x=0$. Therefore,
we need to introduce a cutoff scale $x_{\min}$,
 see~\cite{adnan3}. We then arrive at
 \bear
 f_4(x_{\rm min})-f_4(x)=-i\ln \left(\frac{x^2}{x_{\rm
 min}^2}\right)^\kappa\,,
 \ear with
$\kappa={\alpha \xi}/{(4\pi)}$, and as long as we have the
knowledge of the propagator in one gauge, we can transform it to
any other gauge according to the formula
 \bear
 S_F(x;\xi)=S_F(x;0)\,
 {\rm e}^{-i\big(f_4(x_{\min})-f_4(x)\big)}=S_F(x,0)\,
 \Big(\frac{x^2}{x_{\rm min}^2}\Big)^{-\kappa}\,.
 \label{propagator}
 \ear
And again, after Fourier transforming to momentum space, one
obtains higher order information for $F(k;\xi)$ and
$\mathcal{M}(k;\xi)$, whose lowest order values are: $F(k,0)=1$
and $\mathcal{M}(k;0)=0$. It is shown in~\cite{adnan3}
 that the perturbative
expansion of the result in~\cite{adnan3} for the fermion
propagator matches onto its known one-loop results up to gauge
independent terms at that order.

As is clear from Eq. ($\ref{propagator}$), it encapsulates non-perturbative
information about the propagator. One can Taylor
expand $\big(\frac{x^2}{x_{\rm min}^2}\big)^{-\kappa}$ and
substitute it back into Eq.~(\ref{propagator}) to arrive at:
 \bear
 S_F(x;\xi)=S_F(x;0)\,\Big(1-\log[\frac{x^2}{x_{\rm
  min}^2}]
  \kappa+\frac{1}{2}\log[\frac{x^2}{x_{\rm
   min}^2}]^2\kappa^2+\cdots \Big)\,. \label{lkf-taylor-ex}
  \ear

     It has been argued in~\cite{LKFT-Loops} that this expansion
reproduces correct leading logarithms to any arbitrary order in
perturbation theory. In~\cite{LK,fradkin}, a formula for the
behavior of the 3-vertex under changes of the gauge parameter was
also obtained, although only in coordinate space. However, its
consequences for momentum space calculations have never been
explored, and it is not straightforward to do so. This 
LKFT for the vertex is much easier to study in the worldline
formalism, which is what we take up in section~\ref{LKFgen}. In
order to do this, we start developing the formalism in the next
section.

\section{Multiphoton amplitudes in scalar QED}
\label{multi-photon}
\renewcommand{\theequation}{4.\arabic{equation}}
\setcounter{equation}{0}

\begin{figure}[h]
   \centering
    \includegraphics[width=0.7\textwidth]{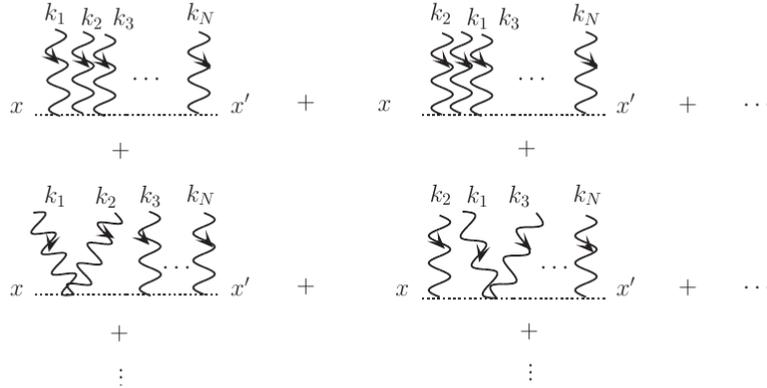}
    \caption{Diagrams contributing to the $N$ - propagator.}
      \label{fig-Npropagator}
 \end{figure}

We will now study the amplitude for the scalar to propagate in
$x$-space from point $x'$ to point $x$, and to absorb or emit $N$
photons along the way  with fixed momenta and polarizations (this
object was called ``N-propagator'' in \cite{100}). It corresponds
to the diagrams shown in FIG.~\ref{fig-Npropagator}.

According to the above diagrams, the worldline representation of
this amplitude is
\bear
 \Gamma[x,x';k_1,\veps1;\cdots;k_N,\veps
N]&=&(-ie)^N\intT\int_{x(0)=x'}^{x(T)=x}\matD x(\tau)\,\eTx\non
&&\times\int_0^T\prod_{i=1}^Nd\t i V^A_{\rm
scal}[k_1,\veps1]\cdots V^A_{\rm scal}[k_N, \veps N]\,,\non
\label{bk-open}
 \ear
  where the photon vertex operator $V_{\rm
scal}^A[k,\vaeps]$ has been introduced in Eq.~(\ref{defvertop}).
We will now derive a closed-form expression for this amplitude.
For this purpose, it will be convenient to  formally rewrite the
vertex operator as
\bear V_{\rm
scal}^A[k,\vaeps]=\int_0^Td\tau\vaeps\cdot\dot{x}(\tau)\,{\rm
e}^{ik\cdot x (\tau)}=\int_0^Td\tau\, {\rm e}^{ik\cdot
x(\tau)+\vaeps\cdot \dot{x}(\tau)}\Big\vert_{{\rm lin}~ \vaeps}\,.
\ear
 Substituting this vertex operator in Eq.~(\ref{bk-open}), and
applying the split in Eq.~(\ref{split}), one gets
 \bear
\Gamma[x,x';k_1,\veps1;\cdots;k_N,\veps N]&=&(-ie)^N\intT\, {\rm
e}^{-\frac{1}{4T}(x-x')^2}\int_{q(0)=q(T)=0}\matD q(\tau)\,\eTq\non
&\times&\int_0^T\prod_{i=1}^Nd\t i\,{\rm
e}^{\sum_{i=1}^N\big(\veps i\cdot \frac{(x-x')}{T}+\veps
i\cdot\dot{q}(\t i)+ik_i\cdot (x-x')\frac{\t i}{T} +ik_i\cdot
x'+ik_i\cdot q(\t i)\big)}\Big\vert_{{\rm lin}(\veps 1\veps2\cdots
\veps N)}\,. \non \label{master-open-scalar}
 \ear
 After completing the square in the
exponential, we obtain the following tree-level
``Bern-Kosower-type formula'' in configuration space:
 \bear \Gamma[x,x';k_1,\veps1;\cdots;k_N,\veps N]
&=&(-ie)^N\intT\, {\rm e}^{-\frac{1}{4T}(x-x')^2}\big(4\pi
T\big)^{-\frac{D}{2}}\non && \hspace{-3.5cm}
\times\int_0^T\prod_{i=1}^Nd\t i\,{\rm e}^{\sum_{i=1}^N\big(\veps
i\cdot \frac{(x-x')}{T}+ik_i\cdot (x-x')\frac{\t i}{T}+ik_i\cdot
x'\big)}\, {\rm e}^{\sum_{i,j=1}^N\big[\Delta_{ij}\kk
ij-2i\ddel_{ij}\veps i\cdot k_j-\ddeld_{ij}\epseps
ij\big]}\Big\vert_{{\rm lin}(\veps 1\veps2\cdots \veps N)}\,.\non
\label{bk-like-x}
 \ear
Now, we also Fourier transform the scalar legs of the master
formula in Eq.~(\ref{bk-like-x}) to momentum space,
\bear \Gamma[p;p';k_1,\veps1;\cdots;k_N,\veps
N]=\int d^Dx\int d^Dx'\, {\rm e}^{ip\cdot x+ip'\cdot x'}\,\Gamma[x,x';k_1,\veps1;\cdots;k_N,\veps N]\,. \ear

\begin{figure}[h]
   \centering
    \includegraphics[width=0.45\textwidth]{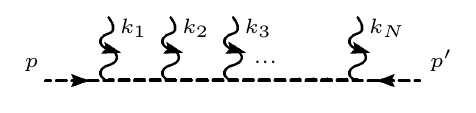}
 \caption{Multi-photon Compton-scattering diagram.}
      \label{fig-compton}
      \end{figure}

This gives a
representation of the multi-photon Compton scattering diagram
as depicted in FIG.~\ref{fig-compton} (together with all the
permuted and ``seagulled'' ones).

\vspace{20pt}

 \vspace{20pt}
      
Changing
the integration variables $x,x'$ to
$$x-x'=x_- ~~~~{\rm and}~~~~ x+x'=2x_+\,,$$
the integral over $x_+$ just produces the
usual energy-momentum conservation factor:

 \bear \Gamma[p;p';k_1,\veps1;\cdots;k_N,\veps N]
&=&(-ie)^N(2\pi)^D\delta^D\Big(p+p'+\sum_{i=1}^N k_i\Big)
\int_0^\infty dT\,{\rm e}^{-m^2T}(4\pi T)^{-\frac{D}{2}}
\int d^Dx_- \,{\rm e}^{-\frac{1}{4T}x_-^2}
\non &&
\hspace{-2cm} \times\int_0^T\prod_{i=1}^N d\t i\, {\rm e}^{ix_-\cdot
\big(p+\sum_{i=1}^N\frac{k_i\t i}{T}\big)}\,{\rm e}^{\sum_{i=1}^N \frac{\veps i\cdot
x_-}{T}}\, {\rm e}^{\sum_{i,j=1}^N\big[\Delta_{ij}\kk
ij-2i\ddel_{ij}\veps i\cdot k_j-\ddeld_{ij}\epseps
ij\big]}\Big\vert_{{\rm lin}(\veps 1\veps2\cdots \veps N)}\,.\non
\label{e1} \ear

After performing also the $x_-$ integral, and some rearrangements, one arrives
at

%

\bear
&&\Gamma[p;p';k_1,\veps1;\cdots;k_N,\veps
N]=(-ie)^N(2\pi)^D\delta^D\Big(p+ p' +\sum_{i=1}^N
k_i\Big)\int_0^\infty dT\, {\rm e}^{-T(m^2+p^2)}\non
&\times&\int_0^T\prod_{i=1}^{N} d\t i\, {\rm
e}^{\sum_{i=1}^N(-2k_i\cdot p\t i+2i\veps i\cdot
p)+\sum_{i,j=1}^N\big[(\frac{\vert \t i-\t j\vert}{2}-\frac{\t i+\t j}{2})\kk
ij-i({\rm sign}(\t i-\t j)-1)\epsk ij+\delta(\t i-\t
j)\epseps ij\big]}\Big\vert_{{\rm lin}(\veps 1\veps2\cdots \veps
N)}\,. \nonumber\\ \label{master-bk-open}
 \ear
This is our final representation of the $N$ - propagator in
momentum space. It is important to mention that it gives the {\sl
untruncated} propagator, including the final scalar propagators
on both ends. On-shell it corresponds to multi-photon Compton
scattering, while off-shell it can be used for constructing
higher-loop amplitudes by sewing. Since this momentum space
version involves the integration variables only linearly in the
exponent, for any given ordering of the photon legs it is
straightforward to do the integrals and verify that they
correspond to the usual sum of Feynman diagrams. The main point of
the formula (\ref{master-bk-open}) is its ability to combine all
the $N!$ orderings. This may not appear very relevant at tree
level, but when used as a building block for higher-loop
amplitudes, it leads to integral representations for nontrivial
sums of diagrams.  For example, taking two copies of the $N$ -
propagator, pairing off the photons on each side, and connecting them
by free photon propagators, we can construct an integral
representation of the sum of ladder plus crossed-ladder diagrams,
important for the study of scalar bound states in scalar
QED. For the case of scalar field theory, the usefulness of this construction
has been demonstrated in \cite{nietjoPRL,100}. 

Let us also remark that the master formula (\ref{master-bk-open}) can be written even more compactly at the expense of introducing some more notation.
Namely, defining
\bear
K_0 &\equiv& p\, , \nonumber\\
K_i &\equiv& k_i\,\,, \,\, i=1,\ldots ,N\, , \nonumber\\
K_{N+1} &\equiv& p'\,,\non \label{defK} \ear as well as $\tau_0 =
T \, , \tau_{N+1} =0 $ and $\varepsilon_0 = \varepsilon_{N+1} = 0$
the exponent of the master formula can be rewritten with the help
of energy-momentum conservation, such as to arrive at the
following form:
\bear &&\Gamma[p;p';k_1,\veps1;\cdots;k_N,\veps
N]=(-ie)^N(2\pi)^D\delta^D\Big(p+ p' +\sum_{i=1}^N k_i\Big)\int_0^\infty
dT\, {\rm e}^{-m^2T}\non &\times&\int_0^T\prod_{i=1}^{N} d\t i\,
{\rm e}^{\sum_{i,j=0}^{N+1}\big[\frac{1}{2}\vert \t i-\t j\vert
K_i\cdot K_j -i{\rm sign}(\t i-\t j)\varepsilon_i\cdot
K_j+\delta(\t i-\t j)\epseps ij \big]}\Big\vert_{{\rm lin}(\veps
1\veps2\cdots \veps N)}\,. \nonumber\\ \label{master-dss} \ear
This form of the momentum space master formula has been previously
obtained by Daikouji {\em et. al.}~\cite{dashsu} by a direct
comparison with the corresponding Feynman-Schwinger parameter
integrals.

\section{The one-loop propagator and vertex in Feynman gauge}
\label{loop-correction}
\renewcommand{\theequation}{5.\arabic{equation}}
\setcounter{equation}{0}

We will now apply the master formula ({\ref{master-bk-open}) to a rederivation of the
one-loop scalar propagator and vertex, at first in Feynman gauge.
Usually such worldline master formulas are used for a direct calculation in parameter space,
and the fact that part or all of the momentum integrals of the corresponding Feynman diagrams
have effectively already been done is an advantage.
Here we will, instead, be satisfied with showing that the master formula correctly reproduces those
Feynman integrals. This will not only provide us with a check, but also allow us to draw on the results of
\cite{adnan-4}.

The one-loop scalar propagator, shown in FIG.
\ref{fig-one_loop_propagator} (there is also a second diagram with
a seagull vertex, which however vanishes in dimensional
regularization), could be obtained either from
Eq.~(\ref{scalar-propagator}), with a single factor of $S_i$, and
Fourier transformation, or from the master formula
(\ref{master-bk-open}) with $N=2$ by sewing off the two photons
legs. We prefer to take the second route here. Let us do this
first in Feynman gauge.

\begin{figure}[h]
  \centering
    \includegraphics[width=0.35\textwidth]{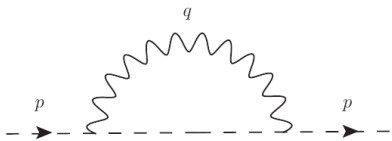}
%
   \caption{One-loop correction to the scalar propagator.}
      \label{fig-one_loop_propagator}
      \end{figure}

\noindent The sewing is done through setting
 \bear
\varepsilon_1^\mu\varepsilon_2^\nu\rightarrow
\frac{\delta_{\mu\nu}}{q^2}\,, \label{feynman gauge} \ear where
$k_1=q,k_2=-q$, and the photon momentum $q$ is integrated over.
After the rescaling $\tau_i = T u_i$, this yields

 \bear \Gamma_{\rm Feyn}^{\rm no-trunc}(p)&=&e^2\int_0^\infty
dT T^2\, {\rm e}^{-T(m^2+p^2)}\int_0^1du_1\int_0^{u_1}du_2\int
\frac{d^Dq}{(2\pi)^D}\non &\times& \Big[(2p+q)^\mu (2p+q)^\nu-  \frac{2}{T}\delta(u_1-u_2)\delta_{\mu\nu}\Big]\frac{\delta^{\mu\nu}}{q^2}\, {\rm
e}^{-T(u_1-u_2)(q^2+2p\cdot q)}\,.\non
\ear
The superscript ``no-trunc'' refers to the above-mentioned fact that this expression still includes the two
external propagators. The term with the delta function corresponds
to the seagull diagram and can be omitted. The parameter integrals
give
\bear
 \int_0^\infty dT T^2\, {\rm e}^{-T(m^2+p^2)}\int_0^1du_1\int_0^{u_1}du_2\,
{\rm e}^{-T(u_1-u_2)(q^2+2p\cdot q)} = \frac{1}{(m^2+p^2)^2[(p+q)^2+m^2]}\,.
\ear
Thus we get
\bear
\Gamma_{\rm Feyn}^{\rm no-trunc}(p)
&=&e^2 \frac{1}{(p^2+m^2)^2}\int \frac{d^Dq}{(2\pi)^D} \frac{(2p+q)^2}{q^2[(p+q)^2+m^2]}\, .\non 
\nonumber\\
\label{fppre}
\ear
The integral can be done using the list of integrals given in Appendix \ref{app-integrals}, leading to
 
\bear
\Gamma_{\rm Feyn}^{\rm no-trunc}(p)
&=& \frac{1}{(p^2+m^2)^2}\frac{e^2}{(4\pi)^{\frac{D}{2}}}(m^2)^{\frac{D}{2}-1}
\Gamma\Big(1-\frac{D}{2}\Big)\Big[2\frac{(m^2-p^2)}{m^2}\,_2F_1\Big(2-\frac{D}{2},1;\frac{D}{2};-\frac{p^2}{m^2}\Big)-1\Big]\,.
\nonumber\\
\label{fp}
 \ear
This agrees with the result in~\cite{adnan-4} after continuation
to Minkowski space, and removal of the external propagators.

Now let us look at the scalar-photon vertex. It can be obtained
from Eq.~(\ref{master-bk-open}) with $N=3$ and the standard
ordering ($\tau_1\geq \tau_2\geq \tau_3$) by sewing photon $1$ and
photon $3$. Our interest is in the form factor decomposition of
the 1PI vertex. Diagrammatically, the 1PI vertex is given by the
three diagrams $a$, $b$ and $c$ depicted in FIG.~\ref{fig-1PI}.
\begin{figure}[h]
  \centering
    \includegraphics[width=0.85\textwidth]{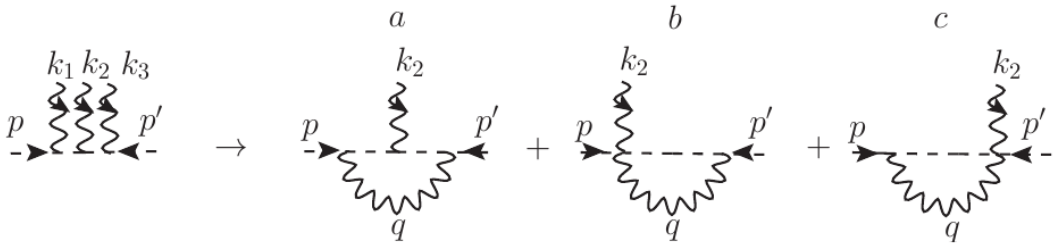}
 \caption{One-loop 1PI vertex from the $N=3$ master formula.}
 \label{fig-1PI}
\end{figure}
Our sewing procedure (still using Feynman gauge) generates these diagrams in the form
 \bear
\Gamma_{\rm vertex}[p,p';k_2,\veps 2]
&=&
\Gamma_a[p,p';k_2,\veps
2]+\Gamma_b[p,p';k_2,\veps 2]+\Gamma_c[p,p';k_2,\veps 2]\non &=&
-e^3(m^2+p^2)(m^2+p'^2)\int_0^\infty dT\,{\rm e}^{-T(m^2+p^2)}
\int_0^Td\t1\int_0^{\t1}d\t2\int_0^{\t2}d\t3
\non && \times \int
\frac{d^Dq}{(2\pi)^D}\bigg\{\frac{ (l_1\cdot  l_3)(l_2\cdot
\veps2)}{q^2} -\frac{( l_3\cdot
\veps2)}{q^2}2\delta(\t1-\t2)+\frac{( l_1\cdot
\veps2)}{q^2}2\delta(\t2-\t3)\bigg\}\non && \times\,
{\rm e}^{-(-2q\cdot p+q^2)\t1-(2k_2\cdot p+k_2^2-2q\cdot
k_2)\t2-(-q^2+2q\cdot (p+k_2))\t3}\,,
\nonumber\\
\label{vertexdiags}
 \ear
where ($q = -k_1 = k_3$)

 \bear
 l_1&=&2p-q\,,\non
 l_2&=&k_2+2(p-q)\,,\non
 l_3&=&2p'+q\,. \nonumber\\
 \ear
The first term inside the curly bracket represents the
$a$-diagram, the second term which contains
$\delta(\tau_1-\tau_2)$ corresponds to the $b$-diagram and the
last term to the $c$-diagram. 
The external propagators have already been removed multiplying by a factor of $-(m^2+p^2)(m^2+p'^2)$.
Let us first look at diagram $a$,
 \bear \Gamma_a[p,p';k_2,\veps 2]
&=&
-e^3 (m^2+p^2)(m^2+p'^2)
\int_0^\infty dTT^3 \, {\rm e}^{-T(m^2+p^2)}
\int_0^1du_1\int_0^{u_1}du_2\int_0^{u_2}du_3\non
&&\times
 \int
\frac{d^Dq}{(2\pi)^D}\frac{(l_1\cdot l_3)(l_2\cdot \veps2)}{q^2}
 \,{\rm e}^{-(-2q\cdot
p+q^2)Tu_1-(2k_2\cdot p+k_2^2-2q\cdot k_2)Tu_2-(-q^2+2q\cdot
(p+k_2))Tu_3}\,,\non
 \ear
 where
 \bear
 l_1\cdot  l_3&=&-q^2+2q\cdot(p-p')+4p'\cdot p\,,\non
 l_2\cdot \veps2&=&k_2\cdot \veps2+2(p-q)\cdot \veps2\,.\non
\ear
Performing the $T$ and $u_i$-integrals leads to %
 \bear
\Gamma_a^\mu[p,p';k_2]&\stackrel{u_1>u_2>u_3}{=}&-e^3 \int
\frac{d^Dq}{(2\pi)^D}\frac{ l_1\cdot
l_3}{q^2}l_2^\mu\frac{1}{(m^2+p^2-2q\cdot
p+q^2)[\underbrace{m^2+p^2+q^2+2k_2\cdot p+k_2^2-2q\cdot
(k_2+p)}_{m^2+(k_2+p-q)^2=m^2+(p'+q)^2}]}\,,\non
 \ear
($\Gamma_a \equiv \varepsilon_{2{\mu}}\Gamma_a^{\mu}$).
Note that the $q$-integral can be rewritten as
 \bear
 \int d^Dq\frac{(l_1\cdot
l_3)l_2^\mu}{q^2[m^2+(p-q)^2][m^2+(p'+q)^2]}&=&\int
d^Dq\Big[\frac{-q^2+2q\cdot (p-p')+4p\cdot p'}{q^2
  {[m^2+(p-q)^2][m^2+(p'+q)^2]} }\Big](k_2^\mu-2q^\mu+2p^\mu)\non
&=&-(k_2^\mu+2p^\mu)K^{(0)}+2K_\mu^{(1)}+2(p^\nu-p'^\nu)\big[(k_2^\mu+2p^\mu)
J_\nu^{(1)}-2J_{\mu\nu}^{(2)}\big]\non
&&+4p\cdot p'\big[(k_2^\mu+2p^\mu) J^{(0)}-2J_{\mu}^{(1)}\big]\,,\non
 \ear
where
 \bear
 K^{(0)}&=&\int
d^Dq\frac{1}{[m^2+(p-q)^2][m^2+(q+p')^2]}\,,\non
K^{(1)}_\mu&=&\int
d^Dq\frac{q_\mu}{[m^2+(p-q)^2][m^2+(q+p')^2]}\,,\non
J^{(0)}&=&\int d^Dq\frac{1}{q^2[m^2+(p-q)^2][m^2+(q+p')^2]}\,,\non
J^{(1)}_\mu&=&\int
d^Dq\frac{q_\mu}{q^2[m^2+(p-q)^2][m^2+(q+p')^2]}\,,\non
J^{(2)}_{\mu\nu}&=&\int d^Dq\frac{q_\mu
q_\nu}{q^2[m^2+(p-q)^2][m^2+(q+p')^2]}\,.\non
 \label{vec-int}
 \ear
The final result for diagram $a$ becomes \bear
 \Gamma_a^\mu[p,p';k_2]&=&-\frac{e^3}{(2\pi)^D}\bigg\{(p'^\mu-p^\mu)K^{(0)}+2K_\mu^{(1)}+2(p^\nu-p'^\nu)\big[(p^\mu-p'^\mu)
 J_\nu^{(1)}-2J_{\mu\nu}^{(2)}\big] \non
 && \hspace{0.2cm} + 4(p\cdot p')\big[(p^\mu-p'^\mu) J^{(0)}-2J_{\mu}^{(1)}\big]\bigg\}\,.\non
\ear
The parameter integrals which appear for the calculation of
diagram $b$ are similar to the ones for the scalar propagator.
Let us present here just the final result, 
relegating the details to Appendix \ref{app-details}: 

 \bear
 \Gamma^\mu_b(p')=\frac{1}{2}\frac{e^3m^{D-4} p'^\mu}{(4\pi)^{\frac{D}{2}}}
\Gamma\Big(1-\frac{D}{2}\Big)\bigg\{\Big(\frac{m^2}{p'^2}-3\Big)\,
_2F_1\Big(2-\frac{D}{2},1;\frac{D}{2};-\frac{p'^2}{m^2}\Big)
-\frac{m^{2}}{p'^2}\bigg\}\,.
\nonumber\\
 \ear
Diagram $c$ is obtained from diagram $b$ simply by the
replacement 

\bear
\Gamma^\mu_c=\Gamma^\mu_b(p'\rightarrow -p)\,.
\ear
\no All these results are in agreement with the ones presented
in~\cite{adnan-4}.

\section{Generalization of the Landau-Khalatnikov-Fradkin transformation}
\label{LKFgen}
\renewcommand{\theequation}{6.\arabic{equation}}
\setcounter{equation}{0}

We now come to our central topic, which is the generalization of
the LKFT to arbitrary amplitudes in scalar QED (we consider scalar
QED as a pure gauge theory only, without the induced $\lambda
(\phi^{\ast}\phi)^2$ coupling term).

As we have seen, each external photon will be represented by a
vertex operator, Eq.~(\ref{defvertop}), inserted either on a
scalar loop or a line. A gauge transformation
 $\veps i\rightarrow \veps i+\xi k_i$ applied to the $i$th photon creates the integral
 of a total derivative that collapses to boundary terms:
 \bear
 V_{\rm scal}[\veps i,k_i]=\int_0^T d\tau_i \vaeps_{i\mu}
\dot{x}_i^\mu\, {\rm e}^{ik_i\cdot x(\tau_i)}\rightarrow V_{\rm
scal}[\veps i,k_i]-i\xi\int_0^T
\frac{\partial}{\partial\tau_i}\,{\rm e}^{ik_i\cdot
x(\tau_i)}=V_{\rm scal}[\veps i,k_i]-i\xi\Big({\rm e}^{ik_i\cdot
x}-{\rm e}^{ik_i\cdot x'}\Big)\,.
 \ear
 Thus, for the closed loop,
these boundary terms cancel. For the open line case they remain,
but do not contribute to on-shell matrix elements. This is, of
course, just the QED Ward identity, which we need not consider
here further.

More interesting is the case of a change of gauge for all the
internal photons. Each internal photon is represented by a factor
of $-S_i$ with $S_i$ as given in Eq.~(\ref{defS}). And from
Eq.~(\ref{gauge-D}), we can see that a change in the gauge
parameter $\xi$ by $\Delta \xi$ will change $S_i$ by
\bear
\Delta_{\xi}S_i =
\Delta\xi  \frac{e^2}{32\pi^{\frac{D}{2}}}
\Gamma\Big(\frac{D}{2}-2\Big)
\int_0^Td\tau_1\int_0^Td\tau_2
\frac{\partial}{\partial\tau_1}\frac{\partial}{\partial\tau_2}[(x_1-x_2)^2]^{2-\frac{D}{2}}\,.
\label{deltaSi}
\ear
Since the integrand is a total derivative in both variables,
if the photon at least on one end sits on a closed loop, such as in  FIG. ~\ref{linetoloop},
the result will vanish.
\begin{figure}[h]
  \centering
    \includegraphics[width=0.2\textwidth]{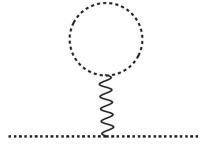}
      \caption{An internal photon with one end on a scalar line and the other one on a loop. }
      \label{linetoloop}
 \end{figure}
Therefore, the gauge transformation properties of an amplitude are
determined by the photons exchanged between two scalar lines, or
along one scalar line. Thus, in the study of the gauge parameter
dependence, we can disregard external photons as well as closed
scalar loops, and it therefore suffices to study the quenched $2n$
scalar amplitude. This amplitude, which we will denote by $A^{\rm
qu}(x_1,\ldots,x_n;x_1',\ldots,x_n'\vert \xi)$, has $n!$
contributions, corresponding to the ways the fields
$\phi(x_1),\ldots,\phi(x_n)$ can be matched with the conjugate
complex fields $\phi^{\ast}(x'_1),\ldots,\phi^{\ast}(x'_n)$:
\bear A^{\rm qu}(x_1,\ldots,x_n;x_1',\ldots,x_n'\vert\xi)=
\sum_{\pi\in S_n} A^{\rm
qu}_{\pi}(x_1,\ldots,x_n;x'_{\pi(1)},\ldots,x'_{\pi(n)}\vert\xi)\,,
 \ear
 where in the partial amplitude $A^{\rm
qu}_{\pi}(x_1,\ldots,x_n;x'_{\pi(1)},\ldots,x'_{\pi(n)}\vert\xi)$
it is understood that the line ending at $x_i$ starts at
$x'_{\pi(i)}$. The worldline representation of this partial
amplitude at the quenched level is ~\cite{feyn-sqed} (compare
Eq.~(\ref{defS}))
\bear
A^{\rm qu}_{\pi}(x_1,\ldots,x_n;x'_{\pi(1)},\ldots,x'_{\pi(n)}\vert\xi) =
\prod_{l=1}^n
\int_0^\infty dT_l\, {\rm e}^{-m^2T_l}\,
  \int_{x_l(0)=x'_{\pi(l)}}^{x_l(T_l)=x_l}\matD x_l(\tau_l)
 \, {\rm e}^{-\sum_{l=1}^n S_0^{(l)} -\sum_{k,l=1}^n S_{i\pi}^{(k,l)}}\,.
\label{Aquenched}
 \ear
 Here, $S_0^{(l)}$ is the free worldline
Lagrangian for the path integral representing line $l$:
\bear
 S_ 0^{(l)} &=&  \int_0^{T_l} d\tau_l \frac{1}{4}\dot{x_l}^2\,,
 \ear
 and $S_{i\pi}^{(k,l)}$ generates all the photons connecting lines $k$ and $l$:
\bear
 S_{i\pi}^{(k,l)} &=& \frac{e^2}{2}
\int_0^{T_k} d\tau_{k}\int_0^{T_l} d\tau_{l}\, \dot{x}^{\mu}_k D_{\mu\nu}(x_k-x_l)\dot{x}^{\nu}_l\,.
\nonumber\\
\label{defSikl}
 \ear
 Thus, after a gauge change,
\bear
A^{\rm qu}_{\pi}(x_1,\ldots,x_n;x'_{\pi(1)},\ldots,x'_{\pi(n)}\vert\xi + \Delta\xi) =
\prod_{l=1}^n
\int_0^\infty dT_l\, {\rm e}^{-m^2T_l}\,
  \int_{x_l(0)=x'_{\pi(l)}}^{x_l(T_l)=x_l}\matD x_l(\tau_l)
 \, {\rm e}^{-\sum_{l=1}^n S_0^{(l)} -\sum_{k,l=1}^n \bigl(S_{i\pi}^{(k,l)}+ \Delta_{\xi}S_{i\pi}^{(k,l)}\bigr)}\,,
 \nonumber\\
\label{Aquenchedtransformed}
 \ear where, from Eq.~(\ref{deltaSi}), we have
 \bear \Delta_\xi  S_{i\pi}^{(k,l)} &=& \Delta\xi
\frac{e^2}{32\pi^{\frac{D}{2}}} \Gamma\Big(\frac{D}{2}-2\Big)
\biggl\lbrace
\bigl[(x_k(T_k)-x_l(T_l))^2\bigr]^{2-D/2}-\bigl[(x_k(T_k)-x_l(0))^2\bigr]^{2-D/2}\non
&&\hspace{90pt}
-\bigl[(x_k(0)-x_l(T_l))^2\bigr]^{2-D/2}+\bigl[(x_k(0)-x_l(0))^2\bigr]^{2-D/2}
\biggr\rbrace \non &=& \Delta\xi  \frac{e^2}{32\pi^{\frac{D}{2}}}
\Gamma\Big(\frac{D}{2}-2\Big) \biggl\lbrace
\bigl[(x_k-x_l)^2\bigr]^{2-D/2}-\bigl[(x_k-x'_{\pi(l)})^2\bigr]^{2-D/2}\non
&&\hspace{90pt}
-\bigl[(x'_{\pi(k)}-x_l)^2\bigr]^{2-D/2}+\bigl[(x'_{\pi(k)}-x'_{\pi(l)})^2\bigr]^{2-D/2}
\biggr\rbrace\,. \non
 \ear Since this depends only on the
endpoints of the scalar trajectories, we can pull the  factors
involving $\Delta\xi$ in Eq.~(\ref{Aquenchedtransformed}) out of
the path integration, leading to
\bear
A^{\rm qu}_{\pi}(x_1,\ldots,x_n;x'_{\pi(1)},\ldots,x'_{\pi(n)}\vert\xi + \Delta\xi) =
T_{\pi}
A^{\rm qu}_{\pi}(x_1,\ldots,x_n;x'_{\pi(1)},\ldots,x'_{\pi(n)}\vert\xi)\,,
 \label{central}
\ear
where
\bear
T_{\pi} &\equiv & \prod_{k,l=1}^N  \, {\rm e}^{- \Delta_{\xi}S_{i\pi}^{(k,l)}}\,. \nonumber\\
\label{defR} \ear
 This is an exact $D$-dimensional result. When
using it in dimensional regularization around  $D=4$, one has to
take into account that the full non-perturbative $A^{\rm qu}$ in
Scalar QED has poles in $\epsilon$ to arbitrary order, so that
also the prefactor $T_{\pi}$, although regular, needs to be kept
to all orders. Here we will consider only the leading constant
term of this prefactor. Thus, we compute
\bear {\rm lim}_{D\to 4} \, {\rm e}^{-
\Delta_{\xi}S_{i\pi}^{(k,l)}} =  \Bigl(r^{(k,l)}_{\pi}\Bigr)^c\,,
\label{lim}
 \ear
 where we have introduced the constant
\bear
c \equiv \Delta\xi  \frac{e^2}{32\pi^2}\,,
\ear
and the conformal cross ratio $ r^{(k,l)}_{\pi}$ associated to the four endpoints of the lines $k$ and $l$,
\bear
 r^{(k,l)}_{\pi}
\equiv \frac{ (x_k-x_l)^2 (x'_{\pi(k)}-x'_{\pi(l)})^2 } {
(x'_{\pi(k)}-x_l)^2 (x_k-x'_{\pi(l)})^2 }\,.
 \ear
 Thus, at the leading order, the prefactor turns into
\bear T_{\pi} = \biggl(\prod_{k,l=1}^N r^{(k,l)}_{\pi}\biggr)^c +
O(\epsilon)\,. \label{leading} \ear We note that for the case of a
single propagator, $s=k=l=1$, Eq.~(\ref{leading})  degenerates into
\bear T =
\biggl\lbrack\frac{(x-x)^2(x'-x')^2}{((x-x')^2)^2}\biggr\rbrack^c\,.
\ear Therefore, if we replace the vanishing numerator
$(x-x)^2(x'-x')^2$ by the cutoff  $(x_{\rm min}^2)^2$, and
$\Delta\xi$ by $\xi$, we recuperate the original LKFT,
Eq.~(\ref{propagator}).

\section{The generalized LKFT in perturbation theory}
\renewcommand{\theequation}{7.\arabic{equation}}
\setcounter{equation}{0}

As in the case of the original LKFT, one would like to know how
the non-perturbative gauge transformation formula in
Eq.~(\ref{central}) works out in perturbation theory. From the
structure of that formula, it is immediately obvious that, given a
calculation of a full $x$ - space amplitude in scalar QED at a
given loop level, it allows one to predict certain higher-order
terms, albeit only gauge-dependent ones. More relevant
from a practical point of view is, however, the change of gauge
parameter at a fixed loop level. How this works diagrammatically
should be clear from the above, but  let us illustrate it with an
example. Consider the twelve-loop contribution to the scalar six-point
function shown in  FIG.~\ref{fig-LKF0}, where it should be
understood that we consider the sum of this diagram together with
all the ones that differ from it only by ``letting photon legs
slide along scalar lines''.

\bigskip

 \begin{figure}[h]
  \centering
    \includegraphics[width=0.3\textwidth]{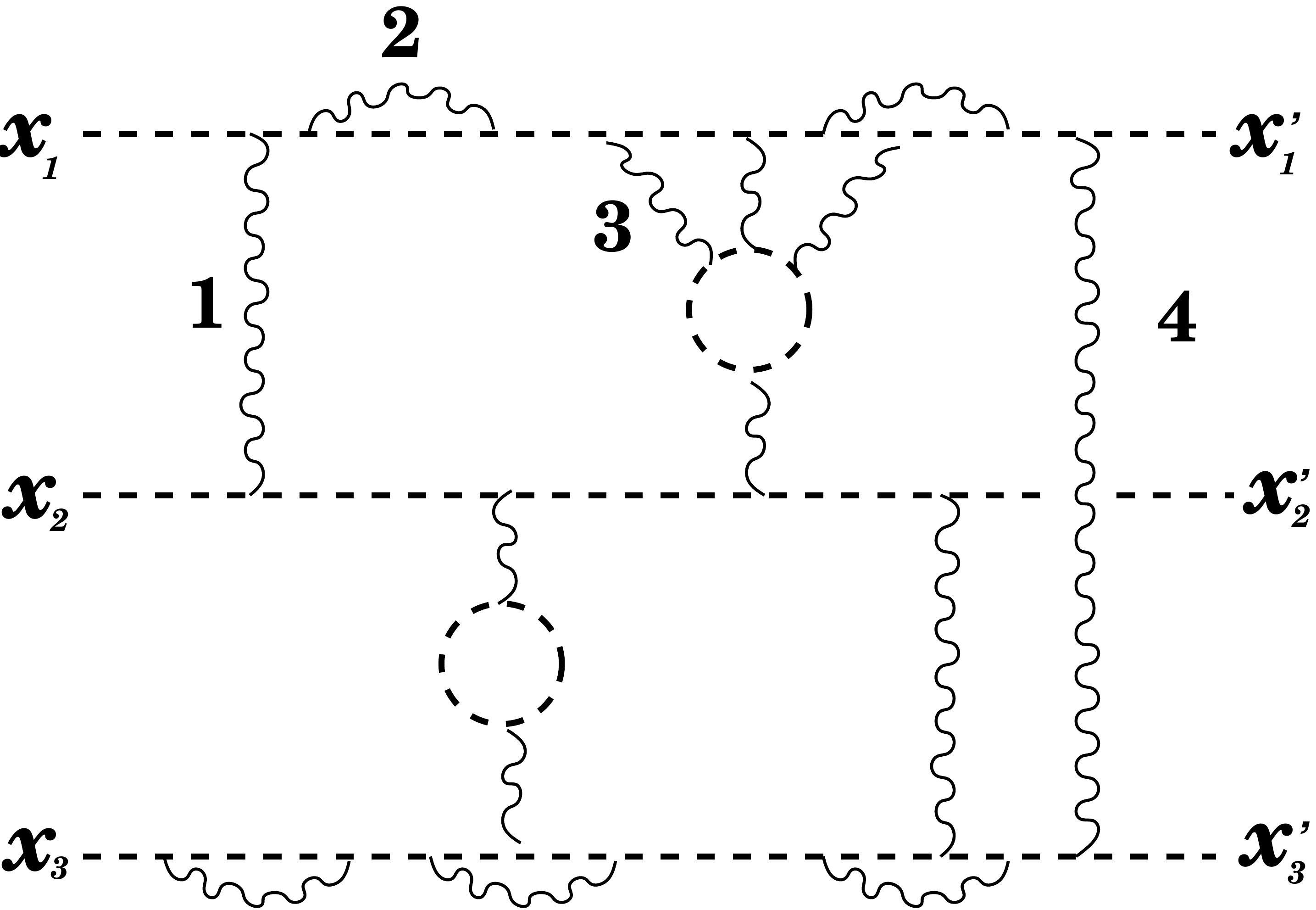}
      \caption{Feynman diagram representing a class of contributions to the six - scalar amplitude at twelve loops.}
      \label{fig-LKF0}
\end{figure}
A gauge parameter change will affect all the photons, except the
ones ending on a loop, and convert a photon connecting lines $k$
and $l$ into a factor $-\Delta_\xi  S_{i\pi}^{(k,l)}$. Thus, the
difference between gauges involves only lower-loop diagrams such
as shown in FIG. \ref{generalLKFT-x}. If the gauge transformation
of the whole set of diagrams is called $\Delta_{\xi}\,{\rm Fig}\,\ref{fig-LKF0}$, we can
write
\bear
 \Delta_{\xi}\,{\rm Fig}\,\ref{fig-LKF0} &=& \big(- 2\Delta_{\xi}S_{i\pi}^{(1,2)})\,{\rm Fig}\,\ref{LKF1}
+\big(- \Delta_{\xi}S_{i\pi}^{(1,1)})\,{\rm Fig}\,\ref{LKF2}
+\big(-2 \Delta_{\xi}S_{i\pi}^{(1,3)})\,{\rm Fig}\,\ref{LKF4}+\cdots \non
&&+\big(-2 \Delta_{\xi}S_{i\pi}^{(1,2)}\big)\big(- \Delta_{\xi}S_{i\pi}^{(1,1)}\big)\,{\rm Fig}\,\ref{LKF12}+\cdots\non
&&+\big(-2 \Delta_{\xi}S_{i\pi}^{(1,2)}\big)\big(- \Delta_{\xi}S_{i\pi}^{(1,1)}\big)\big(-2 \Delta_{\xi}S_{i\pi}^{(1,3)}\big){\rm Fig}\,\ref{LKF124}+\cdots\non &&+\cdots \,.
 \label{KLFT-complicated}
 \ear
Here, on the right-hand side the first line is for gauge
transformation of each photon, one by one. In the second line, we
have simultaneous gauge transformations of all possible pairs of
photons, etc.
 \vspace{20pt}
\begin{figure}[H]
\centering
\subfigure[Gauge transformation of photon $1$.]{%
\includegraphics[width=0.25\textwidth]{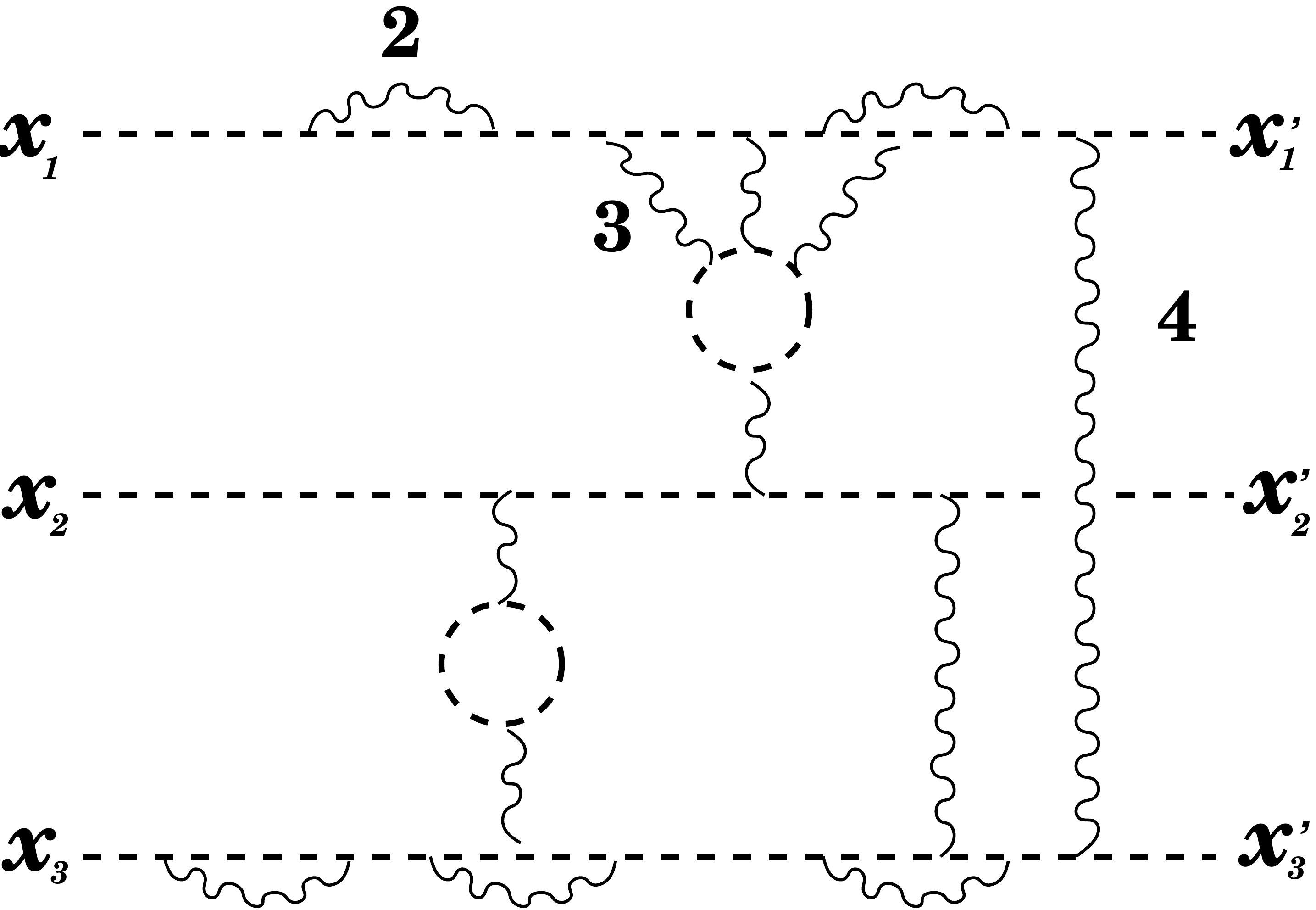}
\label{LKF1}}
\quad
\subfigure[Gauge transformation of photon $2$.]{%
\includegraphics[width=0.25\textwidth]{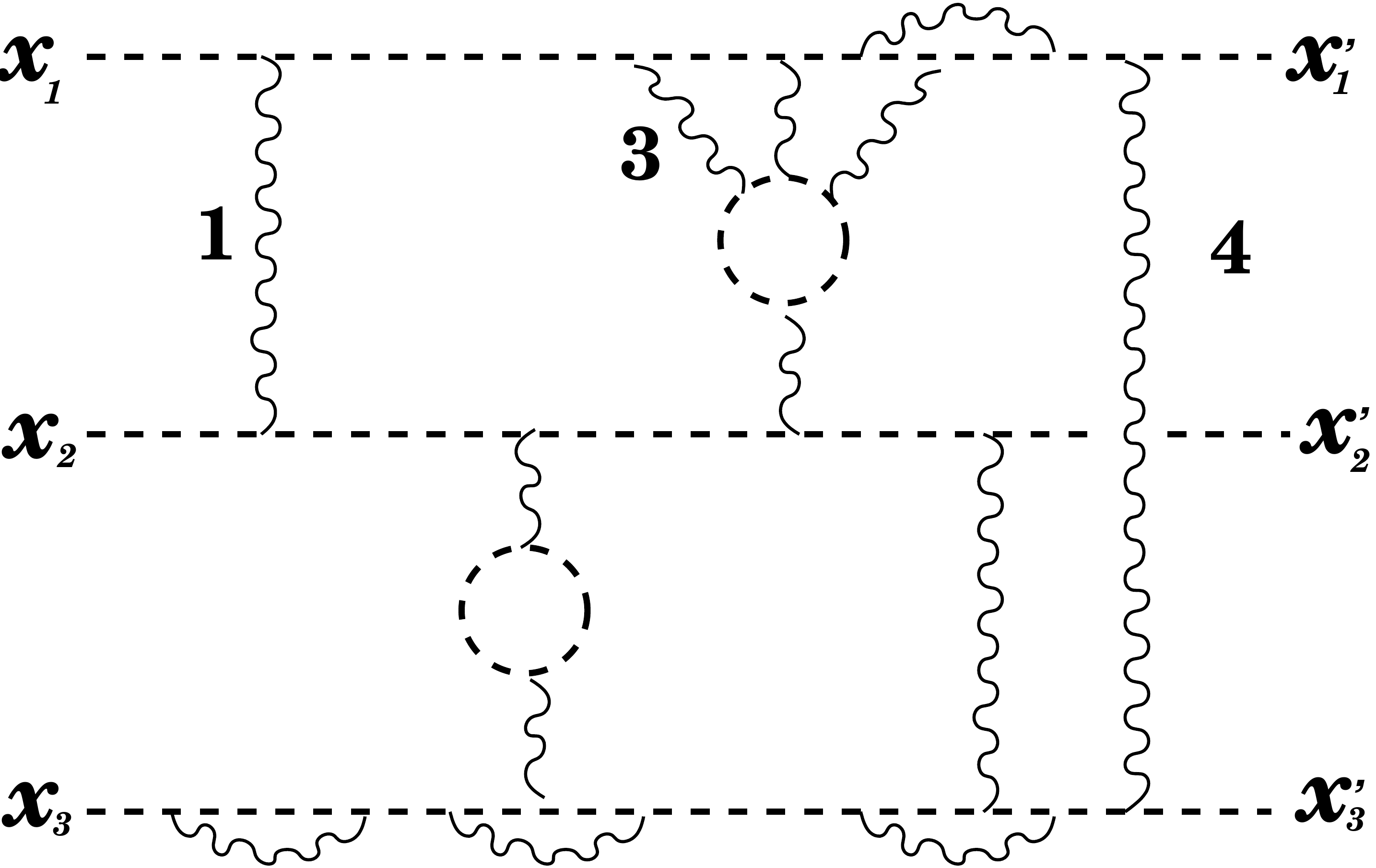}
\label{LKF2}}
\subfigure[Gauge transformation of photon $4$.]{%
\includegraphics[width=0.25\textwidth]{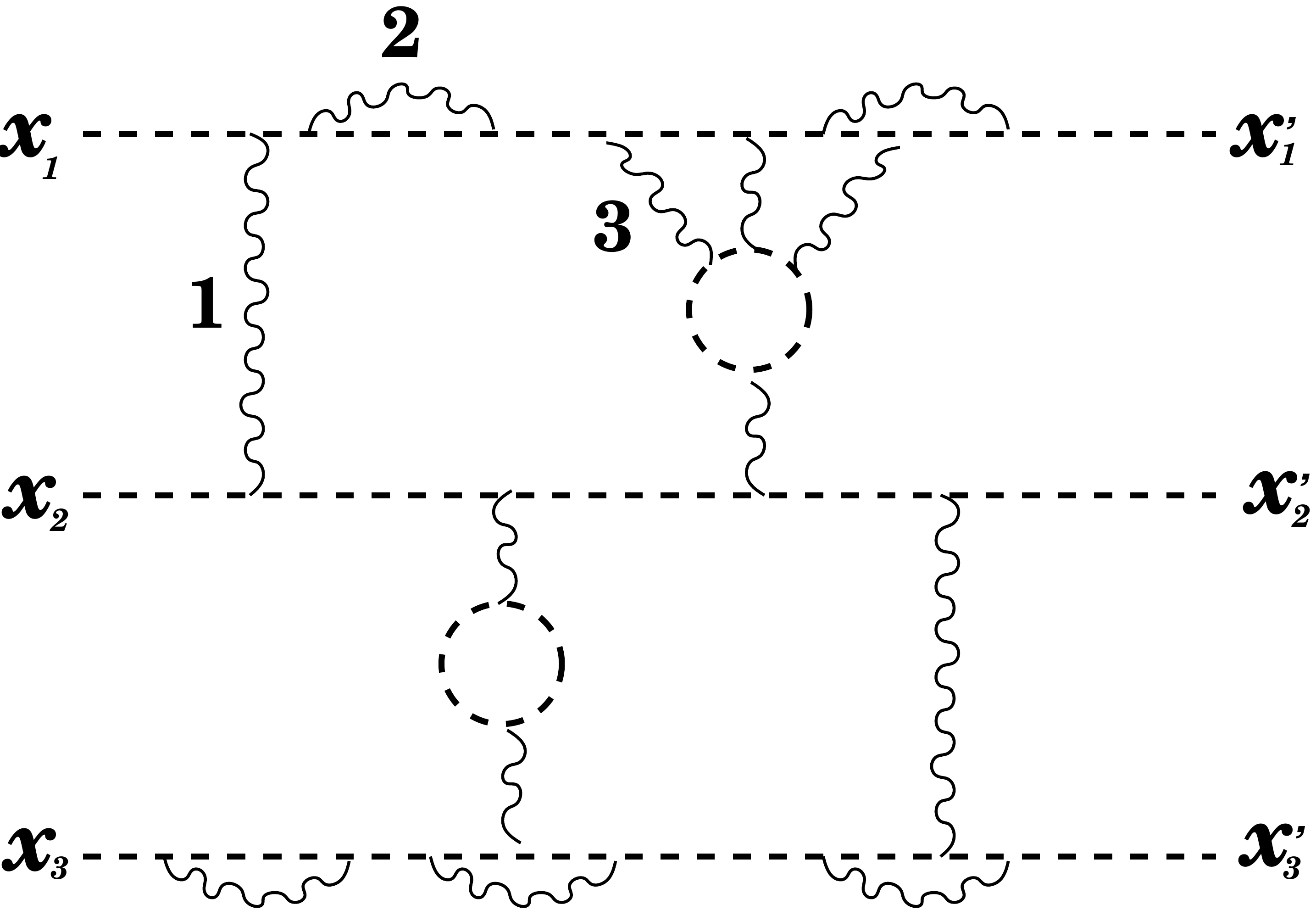}
\label{LKF4}}
\quad
\subfigure[Simultaneous gauge transformation of photons $1$ and $2$.]{%
\includegraphics[width=0.25\textwidth]{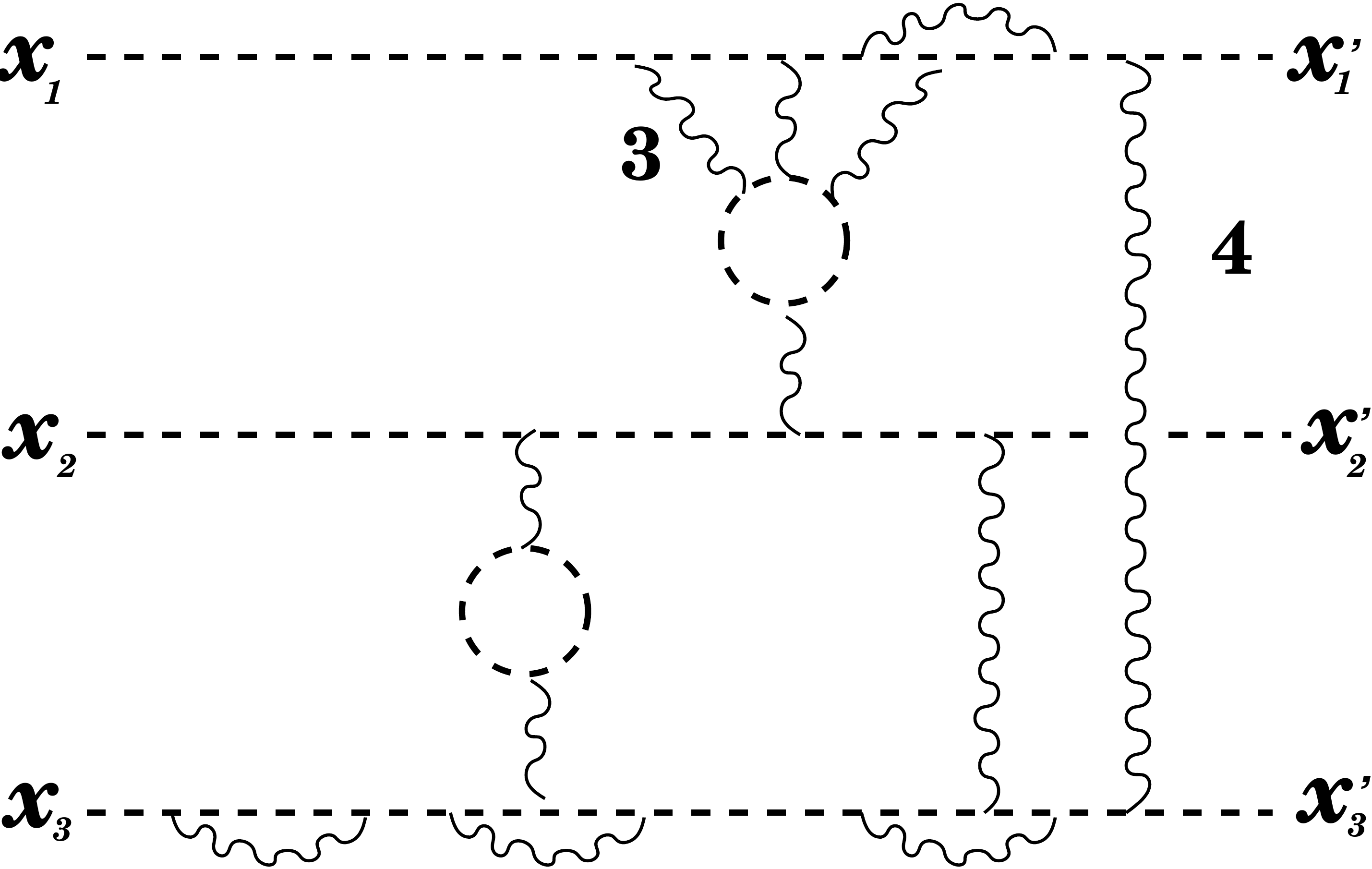}
\label{LKF12}}
\quad
\subfigure[Simultaneous gauge transformation of photons $1$, $2$ and $4$.]{%
\includegraphics[width=0.25\textwidth]{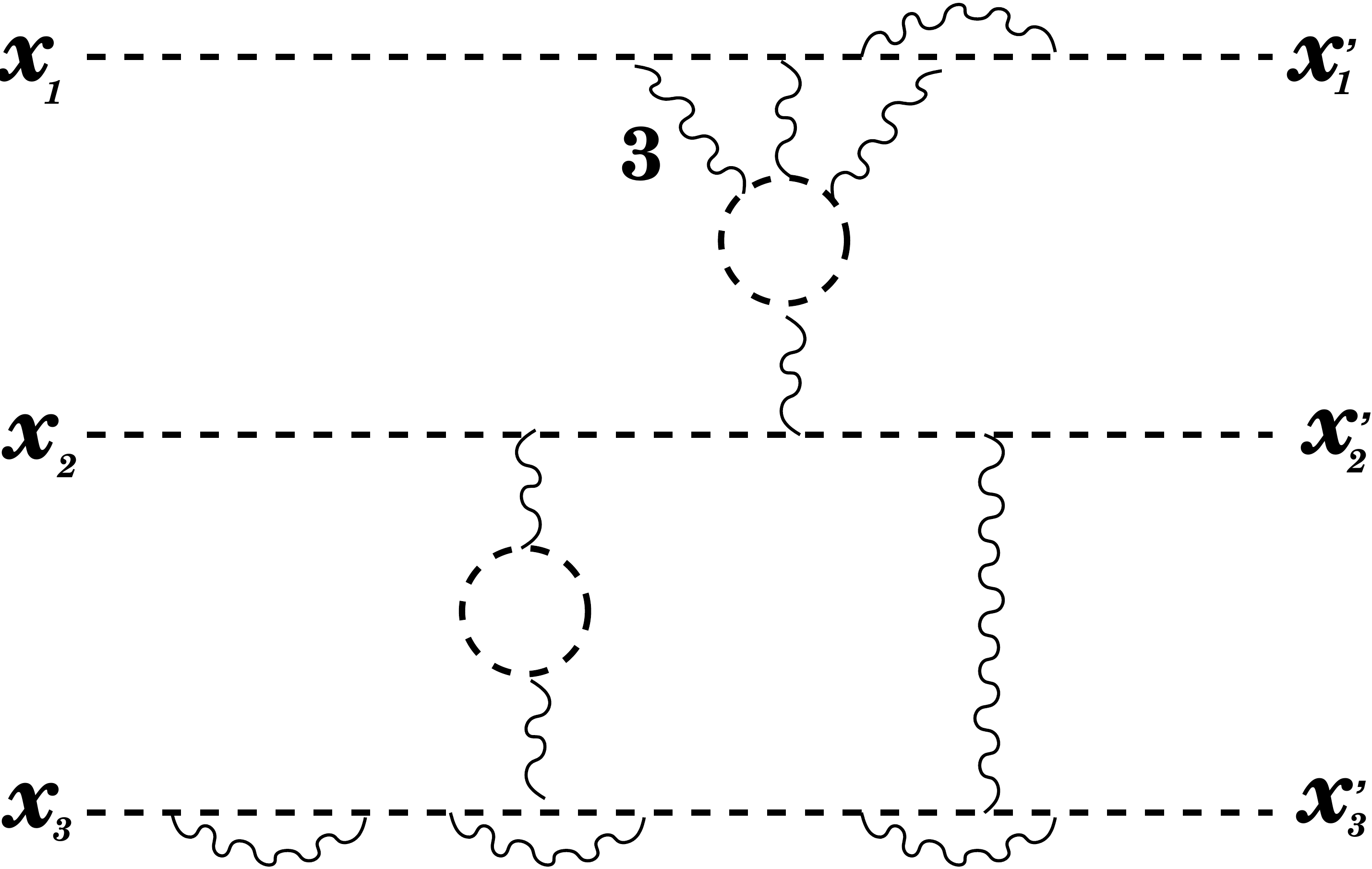}
\label{LKF124}}
\caption{Gauge transformation of internal photons.}
\label{generalLKFT-x}
\end{figure}
 %

\section{The one-loop propagator and vertex in any covariant gauge}
\label{gauge-tr-int-p}
\renewcommand{\theequation}{8.\arabic{equation}}
\setcounter{equation}{0}

As has already been emphasized, mathematically the
LKFT is more difficult to implement in momentum than in
configuration space, and this remains true in our present more
general framework. The non-perturbative transformation formula
(\ref{central}) is not amenable to a direct Fourier
transformation, and neither appears there to be an analogue of the
perturbative formula (\ref{KLFT-complicated}). Nevertheless, the
basic fact that gauge parameter transformations generate only
total derivative terms in the worldline integrals applies, of
course, also in momentum space. This deserves further
investigation; here, we will be satisfied with identifying those
total derivative terms for the case of the one-loop propagator and
vertex, and using them for shifting the above results for Feynman
gauge to an arbitrary covariant gauge in an efficient way.

For this purpose, let us go back to the scalar propagator, FIG. \ref{fig-one_loop_propagator}.
We now use an arbitrary covariant gauge to sew the two photons together:
 \bear
\varepsilon_1^\mu\varepsilon_2^\nu\rightarrow
\frac{\delta^{\mu\nu}q^2-(1-\xi)q^\mu q^\nu}{q^4}\,.
\label{covariant gauge}
 \ear
If we repeat our calculations of section \ref{loop-correction} in an arbitrary covariant
gauge, we get
 \bear
 \Gamma_{\rm propagator}(p)&=&\Gamma_{\rm
Feynman}+\Gamma_\xi=-e^2(m^2+p^2)^2\int_0^\infty dT T^2\, {\rm
e}^{-T(m^2+p^2)}\int_0^1du_1\int_0^{u_1}du_2\int
\frac{d^Dq}{(2\pi)^D}\non &\times& 
(2p^{\mu}+q^{\mu})(2p^{\nu}+q^{\nu})
\Big[\frac{\delta^{\mu\nu}}{q^2}+(\xi-1)\frac{q^\mu
q^\nu}{q^4}\Big]\, {\rm e}^{-T(u_1-u_2)(q^2+2p\cdot q)}\,,\non
\ear
where $\Gamma_{\rm Feynman}$ is the part which
contains $\delta^{\mu\nu}$ and $\Gamma_\xi$ is the gauge part. Note
that the gauge part can be written as a second derivative of the
exponential as \bear
 &&
 (2p^{\mu}+q^{\mu})(2p^{\nu}+q^{\nu})
 (\xi-1)\frac{q^\mu q^\nu}{q^4}\,{\rm e}^{-T(u_1-u_2)(q^2+2p\cdot q)}\non
 &&=(\xi-1)\Big[\frac{2p\cdot q}{q^2}+1\Big]^2\,{\rm e}^{-T(u_1-u_2)(q^2+2p\cdot q)}=-\frac{(\xi-1)}{T^2q^4}\frac{\partial^2}{\partial u_1\partial u_2}\,{\rm e}^{-T(u_1-u_2)(q^2+2p\cdot q)}\,.\non
 \ear
Integrating this over $u_1$ and $u_2$ yields

\bear
\int_0^1du_1\int_0^{u_1}du_2 
\frac{\partial^2}{\partial u_1\partial u_2}\,{\rm e}^{-T(u_1-u_2)(q^2+2p\cdot q)}
=
1- T(q^2+2p\cdot q)- \,{\rm e}^{-T(q^2+2p\cdot q)}
\, .
\ear
The first two terms lead to $q$ - integrals that vanish in dimensional regularization. 
The third one gives a standard scalar two-point integral that is easy to compute,
leading to our following final result for the propagator:
 \bear
 \Gamma_{\rm propagator}(p)=
 \Gamma_{\rm Feynman}+\Gamma_\xi&=&\frac{e^2}{m^2}\Big(\frac{m^2}{4\pi}\Big)^{\frac{D}{2}}
 \Gamma\Big(1-\frac{D}{2}\Big)\bigg\{1-2\frac{(m^2-p^2)}{m^2}\,_2F_1\Big(2-\frac{D}{2},1;\frac{D}{2};-\frac{p^2}{m^2}\Big)\non
&&\hspace{3cm}+(1-\xi)\,
\frac{(m^2+p^2)^2}{m^4}\,_2F_1\Big(3-\frac{D}{2},2;\frac{D}{2};-\frac{p^2}{m^2}\Big)\bigg\}\,.\non
\label{p-s-propagator}
 \ear
%
Now, let us look at the scalar-photon vertex, FIG. \ref{fig-1PI}. In
an arbitrary covariant gauge, (\ref{vertexdiags}) becomes 
 \bear
\Gamma_{\rm vertex}[p,p';k_2,\veps 2]
&=&
\Gamma_a[p,p';k_2,\veps
2]+\Gamma_b[p,p';k_2,\veps 2]+\Gamma_c[p,p';k_2,\veps 2]\non &&
\hspace{-2cm} =
 -e^3(m^2+p'^2)(m^2+p^2)\int_0^\infty
dT\,{\rm e}^{-T(m^2+p^2)}\int_0^Td\t1\int_0^{\t1}d\t2\int_0^{\t2}d\t3\non && \hspace{-2cm} \times\int
\frac{d^Dq}{(2\pi)^D}\bigg\{\Big[\frac{ (l_1\cdot
l_3)}{q^2}-(1-\xi)\frac{ (l_1\cdot q)(l_3\cdot q)}{q^4}\Big]
(l_2\cdot \veps2)\non && \hspace{-2cm}
-2\delta(\t1-\t2)\Big[\frac{
l_3\cdot \veps2}{q^2}-(1-\xi)\frac{( l_3\cdot q)(\veps2\cdot
q)}{q^4}\Big]
+2\delta(\t2-\t3)\Big[\frac{(l_1\cdot \veps2)}{q^2}-(1-\xi)\frac{(
l_1\cdot q)(\veps2\cdot q)}{q^4}\Big]\bigg\}\non && \hspace{-2cm} \times {\rm
e}^{-(-2q\cdot p+q^2)\t1-(2k_2\cdot p+k_2^2-2q\cdot
k_2)\t2-(-q^2+2q\cdot (p+k_2))\t3}\,.\non
\label{amp-vertex}
 \ear
In the same way as for the propagator above,
the gauge parameter dependent part can be obtained from two total
derivatives of the exponential as
 
\bear &&(1-\xi)\frac{l_1\cdot q
l_3\cdot q}{q^4}\Big[{\rm e}^{(2q\cdot p-q^2)Tu_1-(2k_2\cdot
p+k_2^2-2q\cdot k_2)Tu_2+(q^2+2q\cdot p')Tu_3}\Big]\non
&&=(1-\xi)\frac{1}{T^2q^4}\big(\frac{\partial^2}{\partial
u_1\partial u_3}\big)\Big[{\rm e}^{(2q\cdot p-q^2)Tu_1-(2k_2\cdot
p+k_2^2-2q\cdot k_2)Tu_2+(q^2+2q\cdot p')Tu_3}\Big]\,.\nonumber\\
\label{total}
 \ear
Finally, the diagram $a$ in a covariant gauge can be written as (see Appendix \ref {app-details} for the details)

\bear && \Gamma_a^\mu[p,p';k_2] \non &&
=-\frac{e^3}{(2\pi)^D}\Bigg\{(p'^\mu-p^\mu)K^{(0)}+2K_\mu^{(1)}+2(p^\nu-p'^\nu)\big[(p^\mu-p'^\mu)
J_\nu^{(1)}-2J_{\mu\nu}^{(2)}\big]+4p\cdot p'\big[(p^\mu-p'^\mu)
J^{(0)}-2J_{\mu}^{(1)}\big]\non
&&-(\xi-1)(p'^2+m^2)(p^2+m^2)
\bigg\{\biggl\lbrack\frac{\pi^{\frac{D}{2}}(p'^\mu p^2+p^\mu m^2)}{p^2(p'^2+m^2)}\Gamma\big(1-\frac{D}{2}\big)(m^2)^{\frac{D}{2}-3}\,_2F_1\big(3-\frac{D}{2},2;\frac{D}{2};-\frac{p^2}{m^2}\big)-(p\leftrightarrow p')
\biggr\rbrack 
\non
&&\hspace{5cm}-\biggr\lbrack \frac{(\pi)^{\frac{D}{2}}p^\mu}{p^2(p'^2+m^2)}\Gamma\big(1-\frac{D}{2}\big)(m^2)^{\frac{D}{2}-2}\,_2F_1(2-\frac{D}{2},1;\frac{D}{2};-\frac{p^2}{m^2})-(p\leftrightarrow p')\biggr\rbrack
\non
&&\hspace{6cm}+(p^\mu-p'^\mu)I^{(0)}-2I^{(1)}_\mu\bigg\}\Bigg\}\,,\non
\label{diag-a}
 \ear
where, besides the integrals that appeared already in ~(\ref{vec-int}), we need to
evaluate the following ones:
 \bear I^{(0)}&=&\int
d^Dq\frac{1}{q^4[m^2+(p-q)^2][m^2+(q+p')^2]}\,,\non
I^{(1)}_\mu&=&\int
d^Dq\frac{q_\mu}{q^4[m^2+(p-q)^2][m^2+(q+p')^2]}\,.\nonumber\\
\label{defI0I1}
 \ear
Similarly, diagram $b$ yields (see Appendix \ref {app-details}):
 \bear
\Gamma^\mu_b(p')&=&\frac{1}{2}\frac{e^3 m^{D-4}\Gamma\big(1-\frac{D}{2}\big)
p'^\mu}{(4\pi)^{\frac{D}{2}}}\bigg\{\Big(\frac{m^2}{p'^2}-3\Big)\,
_2F_1\Big(2-\frac{D}{2},1;\frac{D}{2};-\frac{p'^2}{m^2}\Big)-\frac{m^{2}}{p'^2}\non
&&\hspace{.7cm}-(\xi-1)\Big(\frac{p'^2+m^2}{p'^2}\Big)\Big[\,_2F_1\Big(2-\frac{D}{2},1;\frac{D}{2};-\frac{p'^2}{m^2}\Big)-\Big(\frac{p'^2+m^2}{m^2}\Big)\,_2F_1\Big(3-\frac{D}{2},2;\frac{D}{2};-\frac{p'^2}{m^2}\Big)\Big]\bigg\}\,.\non
\label{diag-b}
 \ear

Now we can compare our final results with the findings
in~\cite{adnan-4}. For the scalar propagator,
Eq.~(\ref{p-s-propagator})
 is in complete agreement with the
results quoted in~\cite{adnan-4}, after taking into account the
conventions of momentum flow. The same is true for the
scalar-photon 3-point vertex. Notice that in~\cite{adnan-4}, this
result is expressed in terms of nine inequivalent vector and
tensor integrals which are $K^{(0)}, J^{(0)}, I^{(0)},
K^{(1)}_{\mu}, J^{(1)}_{\mu}, I^{(1)}_\mu, J^{(2)}_{\mu\nu},
I^{(2)}_{\mu\nu}$ and $I^{(3)}_{\mu\nu\alpha}$. In our analysis,
the use of total derivative terms has allowed us to reduce the
number of independent integrals by two, i.e., we do not require
$I^{(2)}_{\mu\nu}$ and $I^{(3)}_{\mu\nu\alpha}$ to express the
vertex.

\section{Conclusions}
\label{conclusion}
\renewcommand{\theequation}{9.\arabic{equation}}
\setcounter{equation}{0}

To summarize, in this paper we have applied the string-inspired
worldline formalism to a number of interrelated issues in scalar
QED:

\begin{itemize}

\item
We have rederived the momentum-space Bern-Kosower type master formula
for the tree-level scalar propagator dressed by an arbitrary number of photons,
obtained in \cite{dashsu} by a comparison with Feynman parameter integrals,
starting directly from the worldline path integral representation of this amplitude.
We have also generalized this master formula to the $x$ - space propagator.

\item
We have used the master formula for constructing, by sewing in Feynman gauge, the
one-loop scalar propagator and the one-loop vertex in arbitrary dimension.

\item These momentum-space results were extended to an arbitrary
covariant gauge in a relatively simple way, observing that the
difference terms involve only total derivatives under the
worldline integrals. We have checked that the result agrees with
the earlier calculation presented in~\cite{adnan-4}.

\item In $x$-space, the implementation of changes of the gauge
parameter through total derivatives has allowed us to obtain, in a
very simple way, an explicit non-perturbative formula for the
effect of such a gauge parameter change on an arbitrary amplitude
summed to all loop orders. This formula generalizes the LKFT and
contains it as a special case. At leading order in the $\epsilon$
- expansion it can be written in terms of conformal cross ratios.

\item
We have illustrated with an example how this non-perturbative transformation works
diagrammatically in perturbation theory.

\end{itemize}

All this can be carried through quite analogously for spinor QED.
Although the LKFT for the propagator has the same form in scalar
and spinor QED~\cite{LK}, at higher points difference terms do arise, 
as will be discussed elsewhere~\cite{spin-naoc}. Another
extension of obvious interest is to the non-abelian case. Using
the worldline formalism along the lines of \cite{bbclo-colorful},
a master formula for the scalar propagator dressed by external
gluons has been obtained in \cite{ahbaco-colorful}. As a next
step, this could be used to construct the fully off-shell
quark-gluon vertex and its Ball-Chiu form factor decomposition in
any covariant gauge.

\section{ACKNOWLEDGEMENTS}
We are grateful to O. Corradini, C. Harvey and A.
Ilderton for helpful discussions and correspondence. C. S. thanks
S. Theisen and the Albert-Einstein-Institut Potsdam for
hospitality during the final stage of this work. This research was
supported by: PROMEP, CIC (UMSNH) and CONACyT Grant nos.
DSA/103.5/14/11184, 4.10, CB-2014- 22117 and CB-2014-242461.

\appendix

\section{A list of integrals}
\label{app-integrals}
\numberwithin{equation}{section}
\setcounter{equation}{0}
\bigskip

Here we collect some integrals arising in the calculation of the propagator and vertex 
that permit simple closed-form expressions in terms of the hypergeometric function $\,_2F_1$
in an arbitrary dimension $D$. All of them are easily obtained using standard Feynman-Schwinger parameters. 

\bear
\int\frac{d^Dq}{(2\pi)^D}\frac{1}{m^2+(q+p)^2}&=&\frac{1}{(4\pi)^{\frac{D}{2}}}(m^2)^{\frac{D}{2}-1}\Gamma\Big(1-\frac{D}{2}\Big)\,,\non
\int\frac{d^Dq}{(2\pi)^D}\frac{1}{q^2[m^2+(q+p)^2]}&=&-\frac{1}{(4\pi)^{\frac{D}{2}}}\Gamma\Big(1-\frac{D}{2}\Big)(m^2)^{\frac{D}{2}-2}\,_2F_1\Big(2-\frac{D}{2},1;\frac{D}{2};-\frac{p^2}{m^2}\Big)\,,\non
\int\frac{d^Dq}{(2\pi)^D}\frac{q^{\mu}}{q^2[m^2+(q+p)^2]}&=&
\frac{1}{2}
\frac{\Gamma\Big(1-\frac{D}{2}\Big)(m^2)^{\frac{D}{2}-2}p^\mu}{(4\pi)^{\frac{D}{2}}}
\Big\lbrace \bigl(1+\frac{m^2}{p^2}\bigr)\,_2F_1\Big(2-\frac{D}{2},1;\frac{D}{2};-\frac{p^2}{m^2}\Big) - \frac{m^2}{p^2} \Bigr\rbrace\, , \non
\int \frac{d^Dq}{(2\pi)^D}\frac{1}{q^4[m^2+(q+p)^2]} &=& 
\frac{\Gamma\Big(1-\frac{D}{2}\Big)}{(4\pi)^{\frac{D}{2}}}(m^2)^{\frac{D}{2}-3}\, _2F_1\Big(3-\frac{D}{2},2;\frac{D}{2};-\frac{p^2}{m^2}\Big)\,,\non
\int \frac{d^Dq}{(2\pi)^D}\frac{q^\mu}{q^4[m^2+(q+p)^2]}&=&
\frac{1}{2}\frac{\Gamma\Big(1-\frac{D}{2}\Big)(m^2)^{\frac{D}{2}-2}p^\mu}{(4\pi)^{\frac{D}{2}}p^2}
\Big\{\,_2F_1\Big(2-\frac{D}{2},1;\frac{D}{2};-\frac{p^2}{m^2}\Big)\non
&&\hspace{4cm}-\frac{(m^2+p^2)}{m^2}\,_2F_1\Big(3-\frac{D}{2},2;\frac{D}{2};-\frac{p^2}{m^2}\Big)\Big\}\,.
\non
\label{list}
\ear

\section{Calculation of the vertex}
\label{app-details}
\setcounter{equation}{0}
\bigskip

In this appendix we fill in some of the details of the calculation of the first and second vertex diagrams, FIG. \ref{fig-1PI} a and b. 
From Eqs. (\ref{amp-vertex}) and (\ref{total}),  the gauge part of $\Gamma_a[p,p';k_2,\varepsilon_2]$ can be written as 

\bear
\Gamma^\mu_{a,\xi}[p,p';k_2,\varepsilon_2]&\equiv & -e^3(m^2+p'^2)(m^2+p^2)\int_0^\infty
dT\,{\rm e}^{-T(m^2+p^2)}\int_0^Td\t1\int_0^{\t1}d\t2\int_0^{\t2}d\t3\non
&&\times\int
\frac{d^Dq}{(2\pi)^D}l_2^\mu\Big[-(1-\xi)\frac{ (l_1\cdot q)(l_3\cdot q)}{q^4}\Big]\,{\rm e}^{c_1\t1-c_2\t2+c_3\t3}\non
&=&e^3(1-\xi)(m^2+p'^2)(m^2+p^2)\int \frac{d^Dq}{(2\pi)^D}\,\frac{l_2^\mu}{q^4}\,\int_0^\infty dT T^3\,\e^{-T(m^2+p^2)}\non
&&\times\int_0^1du_1\int_0^{u_1}du_2\int_0^{u_2}du_3\frac{1}{T^2}\frac{\partial^2}{\partial u_1\partial u_3}\,\e^{T(c_1u_1-c_2u_2+c_3u_3)}\,,\non
\ear
where $l_2 = k_2 + 2(p-q)$ and

\bear
c_1&=&2p\cdot q-q^2\,,\non
c_2&=&2k_2\cdot p+k_2^2-2q\cdot k_2\,,\non
c_3&=&q^2+2q\cdot p'\,.\non
\ear
The calculation of the parameter integrals is straightforward. Using 

\bear
m^2+p^2-c_1&=&m^2+(p-q)^2\,,\non
m^2+p^2+c_2-c_1&=&m^2+(p'+q)^2\,,\non
m^2+p^2+c_2-c_1-c_3&=&m^2+p'^2\, ,\non
\ear
the result can be written as 

\bear
&&\int_0^\infty dT T\,\e^{-T(m^2+p^2)}
\int_0^1du_1\int_0^{u_1}du_2\int_0^{u_2}du_3\,\frac{\partial^2}{\partial u_1\partial u_3}\, \e^{T(c_1u_1-c_2u_2+c_3u_3)}
= \frac{1}{(m^2+p'^2)[m^2+(p-q)^2]}\non
&&\hspace{1cm}-\frac{1}{[m^2+(p-q)^2][m^2+(p'+q)^2]}-\frac{1}{(m^2+p'^2)(m^2+p^2)}+\frac{1}{(m^2+p^2)[m^2+(p'+q)^2]}\,.\non
\ear
Using this result together with (\ref{defI0I1}) and (\ref{list}) 
one gets the final result for diagram a as given in Eq. (\ref{diag-a}).

Now, let us look at the second diagram of FIG. \ref{fig-1PI}. From (\ref{amp-vertex}),

\bear
\Gamma^\mu_b[p,p';k_2,\varepsilon_2]&=&-e^3(m^2+p^2)(m^2+p'^2)\int_0^\infty dTT^2\,\e^{-T(m^2+p^2)}\int_0^1du_1\int_0^{u_1}du_2\,\int_0^{u_2}\,du_3\non
&&\times\delta(u_1-u_2)\int\frac{d^Dq}{(2\pi)^D}\Big\{-\frac{l_3^\mu}{q^2}+(1-\xi)\frac{q^\mu(l_3\cdot q)}{q^4}\Big\}\,\e^{T(c_1u_1-c_2u_2+c_3u_3)}\,,
\ear
where $l_3 = q+2p'$, and in the gauge part one can rewrite

\bear
(1-\xi)\frac{q^\mu(l_3\cdot q)}{q^4}\,\e^{T(c_1u_1-c_2u_2+c_3u_3)}=(1-\xi)\frac{q^\mu}{T q^4}\frac{\partial}{\partial u_3}\,\e^{T(c_1u_1-c_2u_2+c_3u_3)}\,.
\ear
The delta function kills one of the parameter integrals as 

\bear
\int_0^1du_1\int_0^{u_1}du_2\int_0^{u_2}du_3\delta(u_1-u_2)\,\e^{T(c_1u_1-c_2u_2+c_3u_3)}= \frac{1}{2} \int_0^1du_1\int_0^{u_1}du_3\,\e^{T(c_1-c_2)u_1+Tc_3u_3}\,.
\ear
The first term in the curly bracket after evaluating all the parameter integrals leads to

\bear
I_b^\mu&\equiv &\int\frac{d^Dq}{(2\pi)^D}\int_0^\infty dTT^2\,\e^{-T(m^2+p^2)}\Big[-\frac{q^\mu+2p'^\mu}{q^2}\Big]\int_0^1du_1\int_0^{u_1}du_3\,\e^{T(c_1-c_2)u_1+Tc_3u_3}\non
&=&-\frac{1}{(m^2+p^2)(m^2+p'^2)}\int\frac{d^Dq}{(2\pi)^D}\,\frac{q^\mu+2 p'^\mu}{q^2[m^2+(p'+q)^2]}\,.\non
\ear
This implies that $I_b^\mu \sim p'^\mu$, so that $I_b^\mu$ can be reconstructed from $I_b\cdot p'$.  Multiplying both sides by $p'^\mu$ we obtain

\bear
p'\cdot I_b&=&-\frac{1}{(m^2+p^2)(m^2+p'^2)}\int\frac{d^Dq}{(2\pi)^D}\,\frac{q\cdot p'+2p'^2}{q^2[m^2+(p'+q)^2]}\non
&=&-\frac{1}{(m^2+p^2)(m^2+p'^2)}\int\frac{d^Dq}{(2\pi)^D}\Big\{\frac{1}{2}\big[\frac{1}{q^2}-\frac{1}{m^2+(p'+q)^2}-\frac{m^2-3p'^2}{q^2[m^2+(p'+q)^2]}\big]\Big\}\,.\non
\ear
The first integral vanishes, while the second and third one have been given in (\ref{list}). 

Coming to the gauge part of diagram b, here we have to calculate 

\bear
&&(1-\xi)\int\frac{d^Dq}{(2\pi)^D}\int_0^\infty dTT\,\e^{-T(m^2+p^2)}\frac{q^\mu}{q^4}\int_0^1du_1\int_0^{u_1}du_3\,\frac{\partial}{\partial u_3}\,\e^{T(c_1-c_2)u_1+Tc_3u_3}\non
&&=(1-\xi)\int\frac{d^Dq}{(2\pi)^D}\frac{q^\mu}{q^4}\,\bigg[\frac{1}{(m^2+p^2)(m^2+p'^2)}-\frac{1}{(m^2+p^2)[m^2+(p'+q)^2]}\bigg]\,.\non
\ear
Here again the first integral vanishes and the second one was given in (\ref{list}). Putting all this together one gets the final result for this diagram which was presented in Eq. (\ref{diag-b}).

\end{document}